\documentclass[AMA,STIX1COL,natbib]{WileyNJD-v2}
\usepackage{moreverb}
\usepackage{NJDnatbib}

\newtheorem{prop}[theorem]{Proposition}

\numberwithin{equation}{section}

\newcommand{\E}{\mathbb{E}}

\newcommand{\N}{\mathbb{N}}

\newcommand\BibTeX{{\rmfamily B\kern-.05em \textsc{i\kern-.025em b}\kern-.08em
T\kern-.1667em\lower.7ex\hbox{E}\kern-.125emX}}

\articletype{PREPRINT}%

\received{}
\revised{}
\accepted{}
\copyyear{2025}

\begin{document}

\title{Leveraging Relational Evidence: Population Size Estimation on Tree-Structured Data with the Weighted Multiplier Method}

\author[1,2]{Mallory J Flynn}

\author[1]{Paul Gustafson}

\authormark{Mallory J Flynn \textsc{et al}}

\address[1]{\orgdiv{Department of Statistics}, \orgname{University of British Columbia}, \orgaddress{\state{Vancouver, BC}, \country{Canada}}}

\address[2]{\orgname{British Columbia Centre for Disease Control}, \orgaddress{\state{Vancouver, BC}, \country{Canada}}}

\corres{Mallory Flynn, \email{mallory.flynn@stat.ubc.ca}}

\presentaddress{Department of Statistics, Faculty of Science, University of British Columbia, 2207 Main Mall, Vancouver, BC, V6T 1Z4, Canada}

\abstract[Abstract]{Populations of interest are often hidden from data for a variety of reasons, though their magnitude remains important in determining resource allocation and appropriate policy. One popular approach to population size estimation, the multiplier method, is a back-calculation tool requiring only a marginal subpopulation size and an estimate of the proportion belonging to this subgroup. Another approach is to use Bayesian methods, which are inherently well-suited to incorporating multiple data sources. However, both methods have their drawbacks. A framework for applying the multiplier method which combines information from several known subpopulations has not yet been established; Bayesian models, though able to incorporate complex dependencies and various data sources, can be difficult for researchers in less technical fields to design and implement. Increasing data collection and linkage across diverse fields suggests accessible methods of estimating population size with synthesized data are needed. 
We propose an extension to the well-known multiplier method which is applicable to tree-structured data, where multiple subpopulations and corresponding proportions combine to generate a population size estimate via the minimum variance estimator. The methodology and resulting estimates are compared with those from a Bayesian hierarchical model, for both simulated and real world data. Subsequent analysis elucidates which data are key to estimation in each method, and examines robustness and feasibility of this new methodology.}

\keywords{Bayesian modeling; multiplier method; population estimation; tree; overdose; opioid}

\jnlcitation{\cname{%
\author{M.J. Flynn}, and
\author{P. Gustafson}} (\cyear{2025}), 
\ctitle{Leveraging Relational Evidence: Population Size Estimation on Tree-Structured Data with the Weighted Multiplier Method}, \cjournal{Preprint}, \cvol{2025;00:1--26}.}

\maketitle

\footnotetext{\textbf{Abbreviations:} MM, multiplier method; WMM, weighted multiplier method; ATD, acute toxicity death}

\section{Introduction}\label{intro}
Population size estimation is of interest in many disciplines, as key populations may be hidden from data for a variety of reasons. 
Within public health and epidemiology, determining the magnitude of these populations is important to both policy makers allocating resources, as well as clinicians developing prevention, care, and treatment programs.  Factors such as healthcare accessibility, symptom presentation, stigmatization, or marginalization, may contribute to medical conditions or events not being captured by administrative health data, leading in turn to underestimated or misguided resource allocation.

\textit{Capture-recapture (CRC)} studies are popular methods of population size estimation.  In it's most basic form, CRC estimates are based on two independent samples, where individuals who were captured by the first sample are marked so that they may be identified if recaptured in the second sample.  Under the assumption that all members of the population have equal probability of capture in the second sample, a total population estimate can be obtained by dividing the size of the first sample by the proportion of recaptured individuals existing in the second sample \cite{kimaniCRC}.  The original methodology does employ several assumptions, including equal probability of capture, independence of samples, and demographic closure \cite{abdulqMM}.  Extensions have been made in the literature to address these, as many applications violate one or more of these assumptions.  In particular, methods which incorporate multiple-recapture \cite{crcmultrecap,crcmultrecapold}, dependence between capture events \cite{fienbergBayes}, and individual heterogeneity have been developed \cite{crcsanath}.  While these extensions make CRC more suitable to applications involving health administrative data \cite{jonesCRC2014}, further assumptions about individual heterogeneity must be made, and quantifying dependency between sampling occasions is complex to model. 
In public health, organizations providing data which inform a target population often provide healthcare to that same population, commonly making referrals between one another for follow-up or complementary treatments.  Furthermore, a representative sample is not typically achieved, and individuals from the target population are more likely to appear in multiple data sources along the natural progression through various health service departments.  While dependencies between data sources can be accounted for through the use of interaction terms in log-linear models, this method is inadequate in the presence of referrals and can lead to multiple seemingly plausible models generating widely varying estimates \cite{jonesCRC2014}.

The \textit{multiplier method} is a commonly used alternative to CRC which is based upon back-calculation from a known subpopulation. In it's most general form, a service or unique object identifier is distributed to individuals of a target population. To obtain a size estimate of the target population, the count of individuals receiving this service or unique object is then used with an estimate of the proportion of the population receiving it \cite{fearonMM}.  While the data source for the latter must be representative of the population, the subpopulation receiving the service or object need not be random.  Furthermore, while the two data sources must be independent and define the population in the same terms, the multiplier method is often able to be employed using readily available, existing data \cite{abdulqMM}.  The method is widely implemented in public health, where some service or trait may serve as a unique identifier defining a subpopulation \cite{abdulqMM}.  Health administrative data are often suitable sources of subpopulations, and to estimate proportions, either existing surveys can be used, or resource allocation can simply focus on gathering data on service/trait proportions via sample surveys \cite{gustafMM}.  Although the multiplier method makes it possible to exploit existing information at all levels, it is sensitive to the accuracy of both marginal counts and estimates of proportions, and will not produce reliable or robust population size estimates with variable data \cite{hickmanCRC2005}.  Furthermore, back-calculated estimates are generated using the count of a single subpopulation of the target population; in practice, multiple subpopulations with corresponding estimates of proportions may be available, but an optimal method of combining target population size estimates produced by multiple sources is not yet developed.

As an alternative to simpler methods, a fully Bayesian approach can be implemented to estimate population size and provides a framework capable of synthesizing data from multiple sources and combining expert knowledge to make inference on model values and parameters \cite{bayespop, tancrediBayes, gustafMM, kingBayes, winBUGS, fienbergBayes}. 

A wealth of literature exists demonstrating the utility of both the multiplier method and Bayesian modeling to provide accurate population size estimates and corresponding uncertainty.  The multiplier method is commonly used by public health agencies and institutes globally in the estimation of the size of key populations such as populations at higher risk of blood-borne infectious disease transmission (e.g., HIV, hepatitis), including people who inject drugs \cite{antaIDU2010,deangelisIDU2004,khalidIDU2014} or men who have sex with men \cite{birrellHIV2013,richMSM2017,pazbailey2011,khalidIDU2014,johnstonHIV2011}.  Similarly, Bayesian modeling is widely used to inform unknown population sizes in these same applications \cite{heesterbeekreview2015,prevostHCV2015,sweetingIDU2009} and is also used to estimate the impact of intervention strategies or therapies \cite{irvinevarbayes2019,irvinethnkits2018,irvineHIV2018}. 
While Bayesian inference is a powerful means of estimation in a variety of settings, it needs to be tailored to the specific problems at hand and requires non-trivial theoretical and practical knowledge to implement.  When this is infeasible, the simplicity of the multiplier method makes it an attractive alternative; however, in it's current form, the method is not capable of fully leveraging all sources of available data or a known dependency structure.  When multiple subsets have been observed, pairing these with the respective proportion estimates and back-calculating in each case results in multiple estimates of the target population, which may not agree.  Furthermore, an inherent underlying network structure is implied by dependencies between data sources in this case, and it is of interest to leverage this structure by modeling these dependencies and synthesizing available evidence to inform target population size estimates. Many fields outside of public health and epidemiology could also leverage a general framework to determine unknown target population sizes, such as in criminology, or in estimating the number of war-time casualties.

Here we develop and describe a novel methodology which is built on the principles of the multiplier method, making it simpler to understand and implement than Bayesian modeling while still incorporating favourable Bayesian attributes, such as the ability to synthesize numerous sources of data and leverage underlying relationships and dependencies. A general network structure inherent to a variety of settings is introduced, with the methodology constructed to compute the optimal population estimate on this structure. In a companion paper, we describe software packages developed to ease implementation of the novel multiplier-based methodology, as well as a hierarchical Bayesian model adaptable to any general tree structure \cite{flynncomp}. In another companion paper, we use the methodology to estimate the number of opioid overdoses events in the province of British Columbia, Canada, over a specified calendar period \cite{flynnapplication}.

\section{Methods} 
We base a new methodology on the traditional multiplier method.  A methodology underpinned by this technique may be more accessible than the application of fully Bayesian models via Markov Chain Monte Carlo (MCMC) or Hamiltonian Markov Chain (HMC) sampling, or through approximate Bayesian computation methods (e.g., variational inference).  Extending this simple, existing method by incorporating the ability to synthesize a number of sources of available data or expert opinion, as inspired by Bayesian modeling, enables a broader range of applications and integration of all available data.  In addition, we introduce a means of accounting for and measuring error not possible with the traditional multiplier method. A general tree structure, which is applicable to a variety of practical applications, serves as the underlying topology to which the methodology applies.

\subsection{Tree-Structured Data}

Suppose multiple subsets of a target population have known or estimated size.  If subsets are mutually exclusive, a tree can be constructed by defining the root to be the target population and the leaves to be sub-populations of the root.  Additional nodes may exist along the root-to-leaf paths, further describing nested subgroups which may also be partially observed, and encode a relational structure between sources of available data.

Structuring relationships between a target population and known subgroups has been done as a consequence of understanding pathways of care \cite{opioiddata, flynnapplication}, and other applications in public health and epidemiology could similarly construct such networks.  In particular, once a target population has been identified, the members of this population can often be characterized by their progression through various treatment (or non-treatment) pathways \cite{opioiddata}; the root node represents the total target population of interest, each leaf represents the endpoint of a possible trajectory defined structurally by the root-to-leaf path, and directed edges are associated with the movement/descent proportion of the parent node population.
Some nodes are associated with observed counts, while others remain latent. For paths including healthcare treatment, administrative data are often available to inform the counts of subpopulations, defined by tree leaves.  A partially observed tree's structure can then be exploited to synthesize the evidence and generate an optimal estimate of a target root node population, given the constraints of the data.  

\subsection{Overview of Multiplier Methodology}

In the simplest case, suppose a target population with unknown size, $Z$, has a sub-population with known marginal count, $N$.  Additionally, suppose an independent survey estimates the proportion of the target population belonging to this sub-population, $\hat{p}_N$.  Then an estimate $\hat{Z}$ can be obtained by the traditional multiplier method by computing
\begin{equation}\label{eq:mm}
\hat{Z} = N \cdot \frac{1}{\hat{p}_N}.
\end{equation}
The marginal count is assumed exact, with target population size estimates being sensitive not only to these counts, but to the estimates of proportions being used in back-calculation \cite{fearonMM}.  There are several known sources of systematic error in health administrative databases, which differ based on the characteristics of the underlying group and the system of data capture; data are often only partially observed or suffer from other coverage errors, measurement errors may result from data translation, and non-response or processing errors are often present at the level of data capture.  Data used to inform branching probabilities is also often derived from these databases.  Each of these sources of error is known to contribute to bias in health administrative data, but the extent to which they affect any particular database is often difficult or impossible to measure \cite{statscanerror}.  
  
When multiple counts of sub-populations are available, this scenario may admit multiple estimates of the root node.  A method which considers sources of uncertainty while generating a single, synthesized estimate of the target population size could prove optimal.

\subsection{Overview of Bayesian Methodology}
At the core of Bayesian reasoning is the notion that a combination of past knowledge, in addition to observed data, should update probabilities, so that estimates are based on integrated knowledge of both of these components.  In particular, suppose we are interested in the distribution of a parameter or hypothesis, $\theta$, given a set of observed data, $\mathcal{D}$.  By incorporating past knowledge about $\theta$ through a choice of \textit{prior distribution}, $p(\theta)$  (possibly conditional on some set of hyperparameters $\alpha$), an application of Bayes theorem dictates that
\[
p(\theta|\mathcal{D}) = \frac{p(\mathcal{D}|\theta)\cdot p(\theta)}{p(\mathcal{D})},
\] 
where $p(\mathcal{D}|\theta)$ is the likelihood of the observed data, conditional on the parameter $\theta$, and $p(\theta|\mathcal{D})$ is the posterior distribution \cite{robertbayesbook}.  In practice, a closed form of the posterior distribution $p(\theta|\mathcal{D})$ is often not  obtainable, and numerical approximation techniques are used in place of exact solutions.  The marginal likelihood, $p(\mathcal{D})$, is the distribution of the data marginalized over the parameters. For continuous $\theta$, we have
\[
p(\mathcal{D} ) = \int p(\mathcal{D} | \theta) p(\theta) d\theta.
\]
The marginal likelihood is often difficult to compute as the above integral cannot necessarily be solved in closed form; in the discrete case, it may involve summing over infinitely many values $\theta$.  As the denominator of the expression for $p(\theta | \mathcal{D})$, it serves as a normalizing constant which ensures the distribution integrates to 1, and fortunately, it does not depend on $\theta$.  Thus a commonly used representation of the posterior distribution is through the expression 
\begin{equation}\label{bayesthm}
p(\theta | \mathcal{D}) \propto p(\mathcal{D} | \theta) \cdot p(\theta).
\end{equation}

Bayesian methods are often implemented using MCMC sampling, where a Monte Carlo random simulation is performed using a Markov chain to explore the state space.  
Given the suitability of MCMC in approximating high-dimensional integrals, the method is now extensively applied in Bayesian inference, where it has been a cornerstone in the adoption of this methodology.  A Markov chain is constructed with the posterior distribution as the steady-state distribution, a limiting distribution in which the density over states is no longer changing in time \cite{murphy}.  
After specifying the model and a set of initial conditions by processing observations and choosing hyperparameters, MCMC can then be used to obtain posterior distributions on parameters from the model for estimation purposes.  Model construction is often aided by representing the joint distributions of random variables using a \textit{graphical model} - a directed, acyclic graph consisting of a collection of nodes which represent random variables, and directed edges representing a set of conditional dependence assumptions (also often called a \textit{DAG}).  An ordered Markov property exists between nodes of a \textit{DAG} and their parent, such that a random variable represented by any node depends only on the random variable represented by it's immediate parent, and not higher predecessors \cite{murphy}.  

\subsection{Weighted Multiplier Method}
When population flows through a tree-like data structure and multiple counts are available at some of the leaves, then so long as branching probability estimates are available for each root-to-leaf path, this scenario admits multiple estimates of the root node. 
The simple form of the multiplier method, as in equation \eqref{eq:mm}, is easily extended to be used on such paths, which we define formally as follows:  
\begin{definition}[Informative Path]\label{def:informativepath}
	Let $\mathcal{T}_Z(V,E)$ be a tree with nodes $V$ representing populations, root $Z \in V$, and edge set $E$.  Let $\mathcal{L} \subset V$ be the set of leaves of $\mathcal{T}_Z$.  A path $\gamma(Z, L) \subseteq E$, is called \textit{informative} if the following conditions hold:
	\begin{itemize}
		\item[(i)] $L \in \mathcal{L}$,
		\item[(ii)] an estimate $D_L$ exists for the marginal count of the population represented by node $L$, and
		\item[(iii)] there exists an estimate for the branching probability $p_e$, $\forall e \in \gamma(Z,L)$.
	\end{itemize}
	We further define the set $\mathcal{L^*}$ to be 
	\[
	\mathcal{L^*} = \{L \in \mathcal{L}: \gamma(Z,L) \text{ is \textit{informative}}\}.
	\]
\end{definition} 

Assuming we have tree-structured data, we generate a function to calculate a root estimate from each informative path $\gamma$ ending in leaf $L \in \mathcal{L^*}$:
\begin{equation}\label{eq:mm2}
	\hat{\theta}_L = D_L \cdot \prod_{e\in \gamma(Z,L)} \frac{1}{\hat{p}_e}.
\end{equation}
To capture uncertainty in the estimates $\hat{p}_e$, we undertake a sampling procedure which assigns a distribution to $p_e$, producing $K$ estimates $\hat{\theta}_{L,k}$ after sampling $p_e$ in $K$ runs.  Prior surveys or literature estimates can be used directly to inform parameters of these distributions, or expert knowledge can be used subjectively to impose appropriate distributions on branching, in much the same process as when choosing priors in Bayesian modeling.

In particular, consider two nodes, $A$ the parent of $B$, which are connected by branch $e'$.  An estimate of $p_{e'}$ is derived from a survey with sample of size $n_{e'}$ of a population akin to $A$, which observes $x_{e'}$ of those individuals fall into the defined category at node $B$.  We inform branching parameters directly by setting $p_{e'} \sim Beta(x_{e'} + 1, n_{e'}-x_{e'} +1)$; then for branch $e'$, we may sample multiple draws from this distribution and back-calculate to produce multiple estimates of $A$ (denoted $\hat{A}$).  The process is iterated over branches for paths $\gamma$ with length greater than one.  

In general, an estimate of the root node population size is generated for each leaf $L \in \mathcal{L}^*$ at each run $k$:
\begin{equation}\label{eq:MM}
	\hat{\theta}_{L,k} = D_L \cdot \prod_{e \in \gamma(Z,L)} \frac{1}{\hat{p}_{e,k}}.
\end{equation}
A number of approaches can be taken to produce a synthesized estimate of the root node population size over all runs and all leaves.  For example, an estimate of the root for any run $k$, $\hat{\Theta}_k$, can be produced by simply choosing a leaf uniformly at random from $\mathcal{L^*}$. 
After simulating $K$ runs, the mean value of these randomly chosen estimates, $\{\hat{\Theta}_k\}_{k=1}^K$, can then be taken as the population estimate: 
\begin{equation}\label{eq:nhatsampled}
	\hat{Z}_K = \frac{1}{K}\sum_{k=1}^K \hat{\Theta}_k.
\end{equation}
Alternatively, we could adjust this procedure to weigh less variable paths more heavily so that the combined root population size estimate is biased towards paths with greater certainty in the evidence.  One procedure is as follows: at each run $k$ generate $Q$ independent draws of the set of branching probabilities, each admitting a leaf-specific estimate of the root population, $\hat{\theta}_{k_q,L}$.  Normalized weights can be assigned to each path, $w_{L,k}$, which are proportional to the inverse of the sample variance of the $Q$ estimates.  A final run estimate, $\hat{\Theta}_k$, can then be chosen by sampling from a categorical distribution with $I = |\mathcal{L^*}|$ categories using the normalized weights as probability parameters, $X_{k_Q} \sim Cat(w_{1,k}, ...,w_{I,k})$, and then averaging the $Q$ samples from leaf $\hat{X}_{k_Q}=L$:
\begin{equation}\label{eq:thetakq}
	\hat{\Theta}_k = \frac{1}{Q}\sum_{q=1}^Q \sum_{L=1}^I \hat{\theta}_{k_q,L} \cdot \mathbf{1}_{\{\hat{X}_{k_Q} = L\}}.
\end{equation}
Here, $\mathbf{1}_{\{\hat{X}_{k_Q} = L\}}$ denotes the indicator variable taking the value 1 when $\hat{X}_{k_Q} = L$, and 0 otherwise.  This process can then be iterated over all runs $k \in \{1, ..., K\}$, and a final estimate of the root, $\hat{Z}_K$, can again be given by equation \ref{eq:nhatsampled}.
Note that $w_{L,k}$ is short-hand notation for the function
\begin{equation}\label{eq:weightshorthand}
	w_{L,k} \equiv w\left(L, \hat{\sigma}^2_k(\hat{\theta}_{k_Q,L})\right),
\end{equation}
where $\hat{\sigma}_k^2$ is the sample variance of the $Q$ estimates of the root, $\{\hat{\theta}_{k_q,L}\}_{q=1}^Q$, for run $k$.  

Under certain conditions on the weights, $w_{L,k}$, a programmatically simpler approach generates the same expected estimate of the root population size.  After $K$ runs (with $Q=1$ for each $k$), we can similarly generate weights $w_{L}$ as a function of the sample variance of the $K$ runs, and simply let
\begin{equation}\label{eq:nhatweightedsum}
	\hat{Z}_K = \sum_{L\in \mathcal{L}^*} w_{L} \cdot \bar{\theta}_{K,L},
\end{equation}
where $\bar{\theta}_{K,L}$ is the average of estimates $\{\hat{\theta}_{L,k}\}_{k=1}^K$ for a given leaf, $L$.  Each estimate generated from $L$ depends on a marginal count for leaf $L$ and the joint distribution of branches $e \in \gamma(Z,L)$.  Though this method requires only one iteration in each of $K$ runs (that is, $Q=1$), the following proposition demonstrates that the expected root estimate from each method is equivalent if $K,Q$ are sufficiently large.  
\begin{prop}
	Let $\{w_{L,k}\}_{L=1}^I$ be such that $\sum_L w_{L,k} = 1$, $w_{L,k}\geq 0$ $\forall L$, and $w_{L,k} =  w(L, \hat{\sigma}^2_k(\hat{\theta}_{L,q}))$ a continuous function of the sample variance $\hat{\sigma}_k^2$ of $Q$ samples of $\theta_L$ at each run $k \in \{1,...,K\}$, where $K,Q$ constant and $\E| \theta_L| <\infty$.  
	Then for sufficiently large numbers of runs, $K$, and iterations $Q$ within each run, the expected value of the root estimate, $\hat{Z}_{K,Q}$, given by repeat weighted sampling (equation \eqref{eq:nhatsampled}, with $\hat{\Theta}_k$ as in equation \eqref{eq:thetakq}) can be made arbitrarily close to the expected value of the estimate $\hat{Z}_K$, given by weighted sum in equation \eqref{eq:nhatweightedsum}.
\end{prop}
The proof is left to the supplementary material. 

The estimate given by equation \eqref{eq:nhatweightedsum} is referred to as the \textit{weighted multiplier method (WMM)} estimate of $Z$, where $Z$ represents the root node population size of the tree $\mathcal{T}_Z$.  While the previous discussion has considered raw path estimates $\hat{\theta}_L$ of the root population size, the preceding analysis continues to apply if we proceed with weight generation using log-transformed path estimates.  Where the values $\hat{\theta}_L$ are log-transformed, a normality constraint on the distributions $\theta_L$ is defensible under the assumption that the un-transformed path-specific estimators are $LogNormal$, with common mean and path-dependent variance. In addition, log-transformation results in better behaved numerical analysis. The final root population size estimate is then given by exponentiating the estimate $\hat{Z}$ of weighted log-transformed path estimates.

\subsubsection{Sampling and Weights Generation}
\begin{figure}
	\centering
	\includegraphics[width=0.6\linewidth]{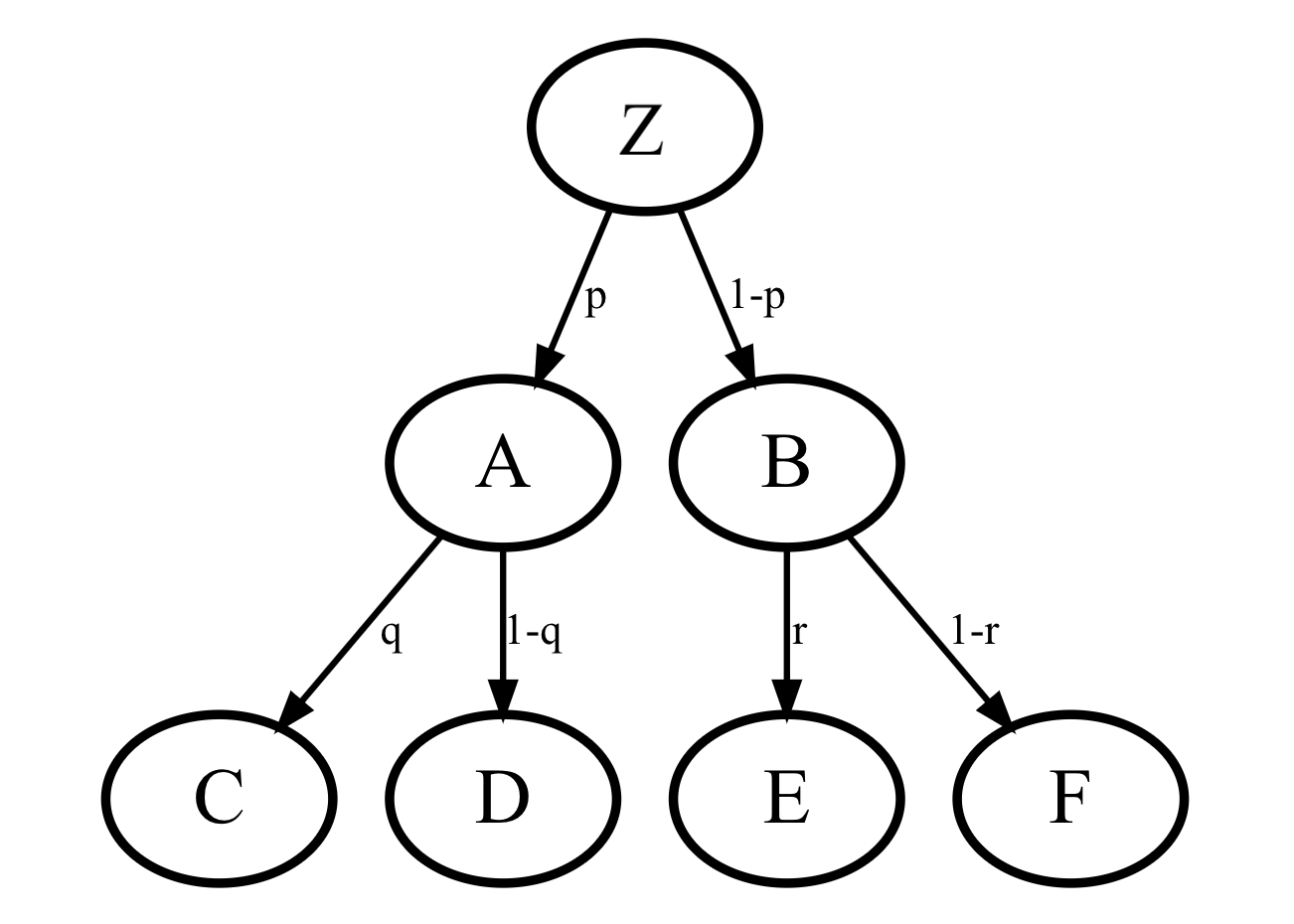}
	\caption{Simple pathways tree.  We assume $C$,$E$ are observed and $D,F$ are latent, while surveys informing $p$, $q$, $r$ are available.}
	\label{fig:babybabytree}
\end{figure}

We illustrate the development of the weight generating procedure by considering a simple tree such as in Figure \ref{fig:babybabytree}, which represents a hypothetical model of relational tree structure representing the flow of data.  Suppose exact marginal counts are available for nodes $C,E$, and estimates of $p,q,r$ may be obtained from three independent surveys performed on different, but related, populations.  In particular, suppose one shows that $x_p/n_p$ individuals move from a population like that at node $Z$ to a subpopulation as defined by node $A$, another survey shows that $x_q/n_q$ individuals move from a population as defined at node $A$ to a subpopulation defined by node $C$, and the third samples $n_r$ individuals, $x_r$ of which move from a population defined as in node $B$ to one defined as in node $E$.  Two estimates, $\hat{\theta}_C(\hat{p},\hat{q})$ and $\hat{\theta}_E(\hat{p},\hat{r})$, of node $Z$ can be obtained using traditional back-calculation methods as in equation \eqref{eq:MM}, using leaves $C$ and $E$.  Branching distributions can be set by choosing $p \sim Beta(x_p+1, n_p-x_p+1)$, $q \sim Beta(x_q+1, n_q-x_q+1)$, and $r \sim Beta(x_r+1, n_r-x_r+1)$, so that the joint distribution $f(p, q, r)$ can be sampled using these distributions.  By Monte Carlo sampling from $f(p,q,r)$, we induce a distribution of $\hat{\theta}_C(p,q)$, $\hat{\theta}_E(p,r)$ which will also give a variance indicating relative uncertainty in each paths' estimate of $Z$, and can be used to determine a weight, $w$. A final estimate of of the target population size, $\hat{Z}$, can be obtained by weighted average:
\[
\hat{Z} = w \hat{\theta}_C(\hat{p}, \hat{q}) + (1-w)\hat{\theta}_E(\hat{p}, \hat{r}). 
\]
Since the two paths are not independent, $\text{Cov}(\hat{\theta}_C, \hat{\theta}_E) \neq 0$. Solving for the minimizing $w$, we find that 
\[ 
w = \frac{\text{Var}\hat{\theta}_E - \text{Cov}(\hat{\theta}_C, \hat{\theta}_E)}{\text{Var}\hat{\theta}_E + \text{Var}\hat{\theta}_C - 2\text{Cov}(\hat{\theta}_C, \hat{\theta}_E)}
\]
minimizes the variance of the weighted estimate, $\hat{Z}$.  

\begin{figure}
	\centering
	\includegraphics[width=\linewidth]{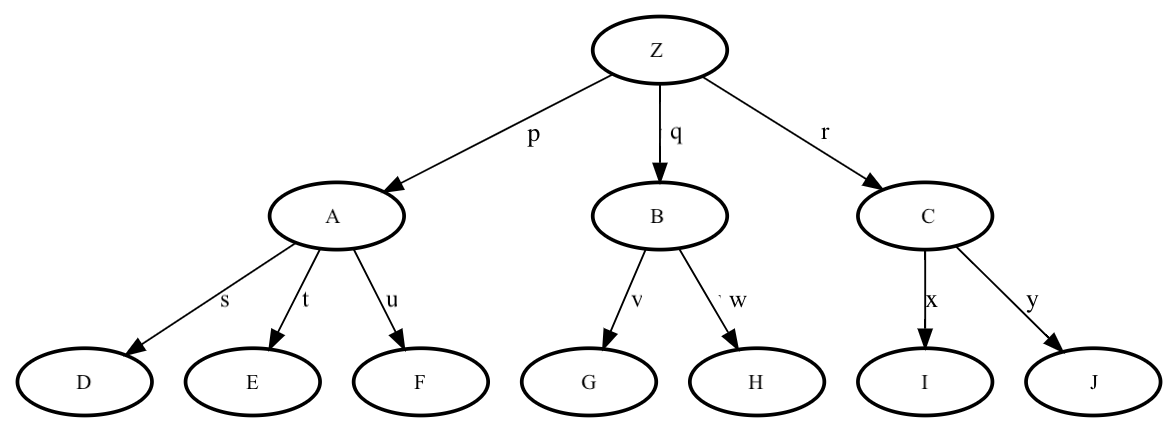}
	\caption{Simple, stylized example of a pathways tree depicting observed or latent nodes.}
	\label{fig:babytree}
\end{figure}

More generally, for $|\mathcal{L^*}|>2$, variance minimizing weights $w_L$ constrained to $\sum_{L \in \mathcal{L}^*} w_L = 1$, can be solved using standard methods, such as Lagrange multipliers.
Variance-minimizing solutions for weights has been solved under a number of assumptions on the general weighted sum of random variables $\theta_s$ \cite{kosoln}, 
\begin{equation}\label{eq:weighted}
    Z = \sum_{s \in \mathcal{S}} w_s  \theta_s,
\end{equation}
under the constraint that
\begin{equation}\label{eq:weightsconstraint}
	\sum_{s \in \mathcal{S}} w_s = 1.
\end{equation} 
A general finding is that setting 
\begin{equation} \label{eq:weights}
	\mathbf{w} = \frac{(\textbf{e}^T \mathbf{\Sigma}^{-1})}{\textbf{e}^T\mathbf{\Sigma}^{-1}\textbf{e}},
\end{equation}
minimizes the variance of $\hat{Z}$, where \textbf{e} is the vector of 1's \cite{kosoln}. To see this, let $\mathbf{\Sigma}$ and $\mathbf{\Sigma}^{-1}$ be the covariance matrix and precision matrix, respectively, of the joint distribution of $\hat{\theta}_L$'s.  Since $\mathbf{\Sigma}$ is positive semi-definite, there exists a matrix $\mathbf{F}$ such that $\mathbf{\Sigma} = \mathbf{F}\mathbf{F}^T$.  In particular, using the eigendecomposition, $\mathbf{\Sigma} = \mathbf{Q}\Lambda \mathbf{Q}^T$, we can take $\mathbf{F} = \mathbf{Q}\Lambda^{1/2}$.  Since the eigenvalues $\lambda_L$ are non-negative, $\sqrt{\lambda_L}$ exists for each $L$, and
\[
\mathbf{F}\mathbf{F}^T = (\mathbf{Q}\Lambda^{1/2})(\mathbf{Q}\Lambda^{1/2})^T = \mathbf{Q}\Lambda \mathbf{Q}.
\]
By Cauchy-Schwarz, for vectors $\mathbf{x}, \mathbf{y}$, we have
\[
(\mathbf{x}^T\mathbf{y})^2 \leq (\mathbf{x}^T\mathbf{x})(\mathbf{y}^T\mathbf{y}).
\]
Setting $\mathbf{x} = \mathbf{F}^T \mathbf{w}$ and $\mathbf{y} = \mathbf{F}^{-1}\textbf{e}$ and substituting, we have
\begin{align*}
	(\mathbf{w}^T\textbf{e})^2 = (\mathbf{w}^T\mathbf{F}\mathbf{F}^{-1}\textbf{e})^2  &\leq (\mathbf{w}^T\mathbf{F}\mathbf{F}^T \mathbf{w})(\textbf{e}^T (\mathbf{F}^{-1})^T\mathbf{F}^{-1}\textbf{e})\\
	&= (\mathbf{w}^T\mathbf{\Sigma} \mathbf{w})(\textbf{e}^T \mathbf{\Sigma}^{-1}\textbf{e}), 
\end{align*}
thus, since $(\mathbf{w}^T\textbf{e})^2 = 1$, we have 
\[
Var(\hat{Z}) = \mathbf{w}^T\mathbf{\Sigma} \mathbf{w} \geq \frac{1}{\textbf{e}^T \mathbf{\Sigma}^{-1}\textbf{e}},
\]
for weights $\mathbf{w}$ satisfying equation \eqref{eq:weightsconstraint}.
Setting $\mathbf{w}$ as in equation \eqref{eq:weights}, we have
\[
\mathbf{w}^T\mathbf{\Sigma} \mathbf{w} = \frac{\textbf{e}^T \mathbf{I}\mathbf{w}}{\textbf{e}^T \mathbf{\Sigma}^{-1}\textbf{e}},
\]
so that
\[
(\mathbf{w}^T\mathbf{\Sigma} \mathbf{w})(\textbf{e}^T \mathbf{\Sigma}^{-1}\textbf{e}) = (\textbf{e}^T\mathbf{w})^2 = 1,
\]
giving equality in Cauchy Schwarz to the minimum solution.  

Weights used in generating the \textit{WMM} estimate in equation \eqref{eq:nhatweightedsum} are given by equation \eqref{eq:weights}. In practice, $\mathbf{\Sigma}^{-1}$ is calculated using pseudo-inverse methods due to the possibility of singular or near-singular covariance matrices $\mathbf{\Sigma}$, which represent the sample covariance matrix. As mentioned previously, population size estimates $\hat{\theta}_L$ are best represented by log-transforming raw path-estimates, $\rho_{K,L}$, before calculating covariance matrices; that is, $\bar{\theta}_{K,L} = \overline{\log(\rho_{K,L})}$ in equation \ref{eq:nhatweightedsum}, where the right hand side are the means of logged population estimates from leaf $L$ and $\hat{Z}$ is the log-transformed estimate of the root population size. 

Extending Figure \ref{fig:babybabytree} beyond binary branching, we consider the tree in Figure \ref{fig:babytree}.  Suppose marginal counts are observed for nodes $D$ and $E$, as well as estimates of probabilities $p$, $s$, and $t$, obtained from a past survey or literature estimate.  If a single survey informs these probabilities, the natural extension is to sample $s,t,u$ from a Dirichlet distribution with parameters informed by this survey.  
However, a single source examining flow of individuals from $A$ to both $D$ and $E$ may not be available.  For example, one survey may observe $x_s$ individuals from a sample of size $n_s$ move from a population akin to $A$ to one akin to $D$, while the remaining $n_s-x_s$ individuals move to the complement $D^c = E \cup F$.  Another survey may similarly estimate $t$ using $x_t/n_t$, while only informing the complement, $t^c=s+u$.  In this case, using these values to determine the parameters of a Dirichlet distribution is not straightforward, and a number of approaches can be used. 

When all sibling branch information is derived from independent sources, one approach is to approximate the $Dirichlet$ distribution of the sibling group by sampling from independent $Beta$ distributions, ignoring the constraint that a complete subset of sibling branch probabilities should sum to 1 (i.e.,$p \sim Beta(x_p + 1, n_p - x_p + 1)$, $s \sim Beta(x_s + 1, n_{s} - x_s +1)$, and $t \sim Beta(x_t + 1, n_{t} - x_t+1)$). The extent to which this approximation affects estimation is explored in simulation, where we compare this independent sampling scheme to a ``mixed sampling'' approach with the WMM which is constructed to handle a range of mixed evidence scenarios, which generates samples which satisfy the constraint on the joint distribution of sibling branches.  These scenarios include the case where all sibling branch knowledge is derived from a single source, where subsets of branches are informed by a mix of independent sources, and lastly, where all observed sibling branches are informed by independent sources, as in the example above.  In the ``mixed sampling'' construction, we use $Dirichlet$ sampling where possible, and rejection schemes and  \textit{importance sampling} where necessary.  This approach is fundamentally supported by applying Bayesian principles as a pre-processing step.  By assuming the true branching distributions of sibling groups are $Dirichlet$, we may consider survey data to have been generated conditional on the true underlying distribution, and set the parameters of the WMM branch distributions to reflect those given by the posterior branching distribution.  The computational realization of these distributions may be accomplished via importance and rejection sampling \cite{flynncomp}.

\subsubsection{General Implementation on Trees}\label{sec:wmmimptrees}

Let $\mathcal{T} \equiv \mathcal{T}_Z(V, E)$ be the tree structure through which data flows, and on which we wish to make inference.  Let $V(\mathcal{T})$ and $E(\mathcal{T})$ be the sets of nodes, $v$, and edges, $e$, in $\mathcal{T}$, respectively, where an edge $e$ is an ordered pair of nodes.  We define a path between two nodes $v_0$ to $v_K$, to be a sequence of edges $\gamma(v_0, v_K) \subseteq E(\mathcal{T})$ connecting $v_0$ to $v_K$.
We further define a substructure, $\mathcal{T}^D \equiv \mathcal{T}_Z^D(V^D, E^D) \subseteq \mathcal{T}$, which represents those nodes and edges of $\mathcal{T}$ for which we have data, so that all nodes and edges in $\mathcal{T} \setminus \mathcal{T}^D$ are latent.

Define $\mathcal{L^*} \subseteq V^D(\mathcal{T})$ as in Definition \ref{def:informativepath}, and let $\mathcal{C}^D(\mathcal{T})$ denote the set of all combinations of branch estimates available to inform $E^D(\mathcal{T})$.
For each $C \in \mathcal{C}^D(\mathcal{T})$, we sample $M$ sets of branching probabilities from distributions determined by the prior knowledge of edges $e \in E^D(\mathcal{T})$.  Each set of samples, indexed by $m$, is combined with marginal leaf counts of each $L \in \mathcal{L^*}$, generating $M$ back-calculated values of the root node per leaf $L$ (not necessarily unique).  This process generates an $M$ by $|\mathcal{L^*}|$ matrix, $\mathbf{M}$, of estimates of the root population size.  In particular, each column of $\mathbf{M}$ represents a leaf, $L$, with observed count, and each row, $m$, corresponds to one sampled realization of the tree, so that a matrix value $M_{m',L'}$ is the back-calculated estimate of the root population given the count at $L'$ and the subset of relevant path probabilities sampled on run $m'$ which are required to perform the back-calculation.  

Depending on the path distributions (i.e., the joint distribution of branching probabilities of edges, $e$, in a path $\gamma$), variation within columns of $\mathbf{M}$ can be large with root estimates within each row differ considerably.  Alternatively, in an ideal setting, estimates of the root population generated by informative paths will largely agree and variation across rows will be low, so that $\mathbf{M}$ may be near-singular.  To generate weights, the inverse of the covariance matrix of $\mathbf{M}$ is required, as in equation \eqref{eq:weights}, but the previous scenarios can result in numerical instability in the former or problems with matrix inversion in the latter.
Singularity can be addressed through a diagonal damping coefficient or through use of a pseudo-inverse.  To help with numerical stability, we log-transform population values in $\mathbf{M}$ before calculating the covariance matrix.  This stabilizes the calculation of the weights, $\mathbf{w} \in \mathbb{R}^{|\mathcal{L^*}|}$, where we impose the constraint $\sum_{L \in \mathcal{L^*}} w_L = 1$.  We can then calculate a root population size estimate, 
\[
\hat{\theta} = f(\mathbf{L} \cdot \mathbf{w}),
\]
where $\mathbf{L}=\log\mathbf{M}$ is the element-wise transformation of $\mathbf{M}$, $(\mathbf{L} \cdot \mathbf{w}) \in \mathbb{R}^{M \times 1}$, and $f$ represents the uniform average.  To obtain the final estimate, we set $\hat{Z} = \exp(\hat{\theta})$. 

Under log-transformation, weights $\mathbf{w}$ are generated using $\mathbf{L}$ and $\hat{Z}$ is a multiplicative function of path estimates raised to the power of the weights $w_i$.  In particular, we have
\begin{align*} 
	\hat{\theta} &=  \sum_m  \left[ \sum_L \frac{\mathbf{w}_L}{M} \cdot \log( \mathbf{M}_{m,L})\right]\\
	&= \log\left(\prod_m \prod_L \mathbf{M}_{m,L}^{\mathbf{w}_L/M} \right)
\end{align*}
and
\[
\hat{Z} = \prod_{m=1}^M \prod_{L \in \mathcal{L^*}}  \mathbf{M}_{m,L}^{\mathbf{w}_L/M}.
\]

The above procedure holds for a single set of estimates $C \in \mathcal{C}^D(\mathcal{T})$, which suggests we have only one set of values informing our branching estimates. When $|\mathcal{C}^D(\mathcal{T})|>1$, at least one branch has more than one plausible estimate.  Where more than one previous data source is available to inform any branching estimates, and it may not be immediately clear which estimate should be used. For $|\mathcal{C}^D(\mathcal{T})|>1$, a two-stage weight generating process could instead be used.  For each $C$, we proceed as above up to the point of generating weights $\mathbf{w}_C$, now dependent on the set $C$.  We then generate a vector of estimates, $\hat{\theta}_C \in \mathbb{R}^M$, defined by 
\[
\hat{\theta}_C = \mathbf{L}_C\cdot \mathbf{w}_C.
\]
The process is repeated for each $C \in \mathcal{C}^D(\mathcal{T})$, and estimates are combined to form the matrix $\hat{\Theta} \in \mathbb{R}^{M \times |\mathcal{C}^D(\mathcal{T})|}$.  The covariance matrix of $\hat{\Theta}$ can then be used to generate weights $\mathbf{W}$ which account for total variance among the possible sets, $C$, of branch data.  A final scalar estimate, $\hat{\psi}$, is then given by
\[
\hat{\psi }= f(\hat{\Theta} \cdot \mathbf{W}),
\]
and can be converted to a population estimate at the original scale by setting $\hat{Z} = \exp(\hat{\psi})$.

\subsubsection{On the Subject of Error}

Sampling branch probabilities from $Dirichlet$ distributions at each run, $k$, in place of using fixed probabilities for branching as in the traditional multiplier method provides some measure of relative path uncertainty which assists in the weighting of informative paths, an important component to synthesizing the data to obtain a root population size estimate.  Several sources of uncertainty are not accounted for using this scheme, though this is in part a matter of construction.  By setting $Beta$ parameters directly using past survey results, the sampling procedure will generate variability in root population size estimates as a function of sample size.  Path-specific estimates can then be compared by their variability as it relates to sample size limitations.  Errors in counts $D_L$ or non-representative samples are common sources of uncertainty which are unaccounted for, except by subjective setting of branch distribution parameters, though a quantitative method of doing so has not been developed.  These sources of uncertainty are often known to affect data commonly used with back-calculation, such as health administrative databases.  Errors in counts $D_L$ would be more naturally accounted for through a methodological extension which assigns distributions to node counts; this extension is simple to implement with the WMM described herein, however we focus here on implementation using fixed node counts, in line with traditional multiplier method approaches.  

The inclusion of prior knowledge in the form of branching distributions imparts a Bayesian element to the WMM methodology.  In addition, the root population size estimator itself, being based on a sum of multiplicative terms involving these random variables, is also assigned a distribution rather than assumed to have a fixed value.  Some important differences, however, do exist, which do not qualify it as a fully Bayesian approach.  Bayesian methodology can also deal with uncertainty in node counts, through the assignment of priors to these counts or the inclusion of ``data uncertainty'' nodes representing missed counts, an approach herein.

\section{Simulated Experiment}

\subsection{Model and Behaviour}
\begin{figure}
	\centering
	\includegraphics[height=0.35\textheight]{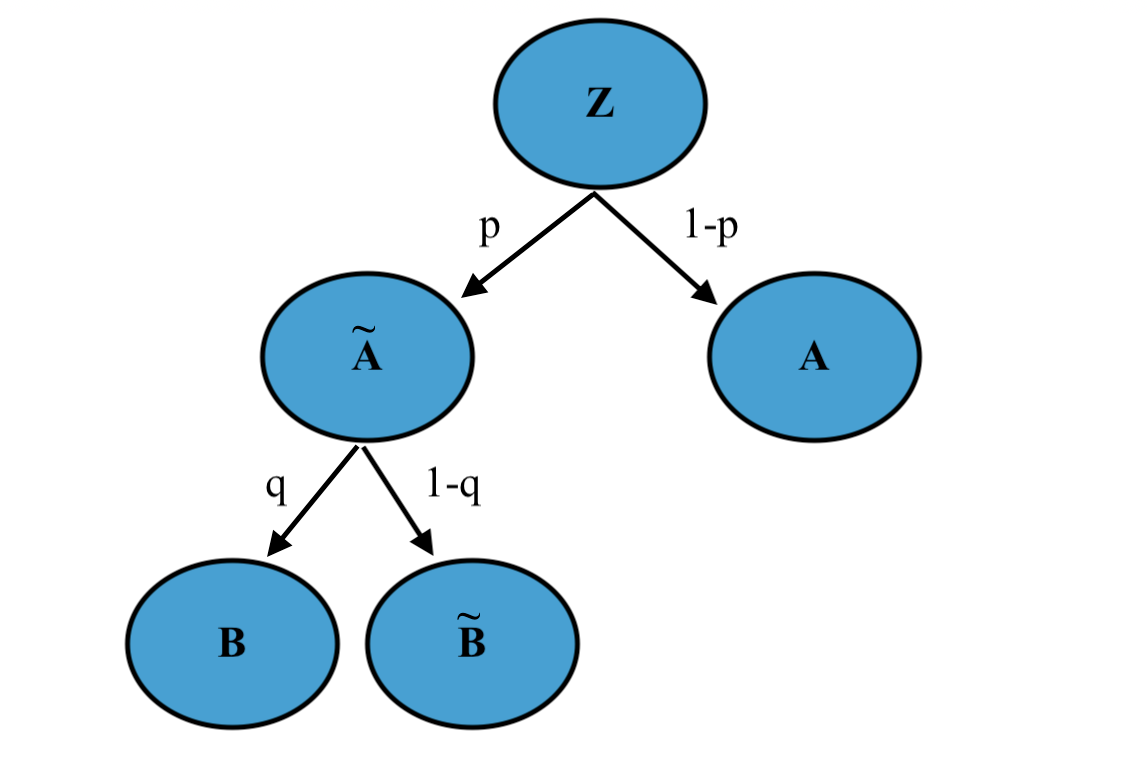}
	\caption{Simple tree used for computation of closed-form posterior distribution and in simulation.}
	\label{fig:simstree}
\end{figure}

Though a closed-form posterior distribution of the target population at the root node may not be achievable for complex tree structures, it is possible to calculate the posterior distribution of the root population size on a simple tree, as in Figure \ref{fig:simstree}, which is used in the simulation modeling.  Having the closed-form expression allows some investigation of how the WMM model behaves in comparison to the Bayesian methods, and how information propagates through the tree.  

Referring to Figure \ref{fig:simstree}, suppose counts for nodes $A,B$ are known, while $\tilde{A}, \tilde{B}$ are latent.  We are interested in examining differences in behaviour of the WMM and the Bayesian model in the limiting cases where (i) we have no information about probability $q$, and (ii) exact information about branching at $q$.  Under scenario (i), having counts at $A$ and $B$ without knowledge of branch $q$ results in only one informative path, $\gamma(Z,A)$, and a straightforward application of the traditional multiplier method would the result in an estimate of $Z$ given by
\[
\hat{Z} = \frac{A}{1-\hat{p}}.
\]  
In case (ii), a single application of the multiplier method again applies, since $A$ is assumed known, so that
\[
\hat{Z} = A + \frac{B}{\hat{q}}.
\]
We now explore the Bayesian model under these limiting scenarios.Let $p \sim Beta(\alpha_p, \beta_p)$ and $q \sim Beta(\alpha_q, \beta_q)$, where $\alpha_p, \alpha_q, \beta_p, \beta_q \in \N$. 
We define constants $C_p$, $C_q$, such that
\begin{align}\label{eq:cp}
	C_p &\equiv C_p(\alpha_p, \beta_p, a, z) \nonumber \\ 
	&= \frac{(\alpha_p)...(\alpha_p+z-a-1)\cdot (\beta_p)...(\beta_p + a - 1)}{(\alpha_p + \beta_p)...(\alpha_p + \beta_p + z-1)}
\end{align}
and
\begin{align}\label{eq:cq}
	C_q &\equiv C_q(\alpha_q, \beta_q, z, a, b) \nonumber \\ 
	&= \frac{(\alpha_q)...(\alpha_q+b-1)\cdot (\beta_q)...(\beta_q + z - a - b-1)}{(\alpha_q + \beta_q)...(\alpha_q + \beta_q + z - a-1)}.
\end{align}
Then by substituting into the integral for 
\begin{equation}\label{eq:integral}
	f(z|a,b) \propto f(z) \int \int f(p) f(q) f(a,b|z,p,q) dp\text{ }dq,
\end{equation}
we have
\begin{equation} \label{eq:simplified}
	f(z | a, b) \propto C_p  C_q  f(z) \cdot  \frac{z!}{a! b! (z-a-b)!}.
\end{equation}
Now, for case (i), we set $\alpha_q, \beta_q = 1$ so that $f(q)$ is flat, which means  
\[
C_q =  \frac{b! (z - a - b)!}{(z - a+1)!},
\]
and we have
\begin{align*}
f(z | a, b) &\propto C_p  f(z) \cdot  \frac{z!}{a!(z - a+1)!}.
\end{align*}
As $C_p$ and $f(z)$ do not depend on $b$, then there is nothing gained from the information at $B$; we do not see, however, the simplified relationship suggested by the multiplier method, in which $\hat{Z} \to A(1-\hat{p})^{-1}$.  

To determine what we may expect if $f(q)$ approaches a point mass as in scenario (ii), we may consider the limit as $\alpha_q, \beta_q \to \infty$ and $\alpha_q/\beta_q \to c$, $c \in (0, \infty)$.  In this case, 
\begin{align}\label{eq:examplecq}
	C_q 
	& \to \frac{1}{c^{z-a-b-1}\cdot (c+1)^{z-a}}\\
	&\equiv c^*(z)
\end{align}
and we have
\[
f(z|a,b) \propto C_p c^*(z) f(z)\cdot \frac{z!}{a!b!(z-a-b)!}.
\]
The above posterior distribution maintains dependence on $f(p)$ through $C_p$, suggesting $f(p)$ is not mathematically irrelevant in the Bayesian model, even when $f(q)$ approaches a point mass.  

\begin{figure}
	\centering
	\includegraphics[height=0.35\textheight]{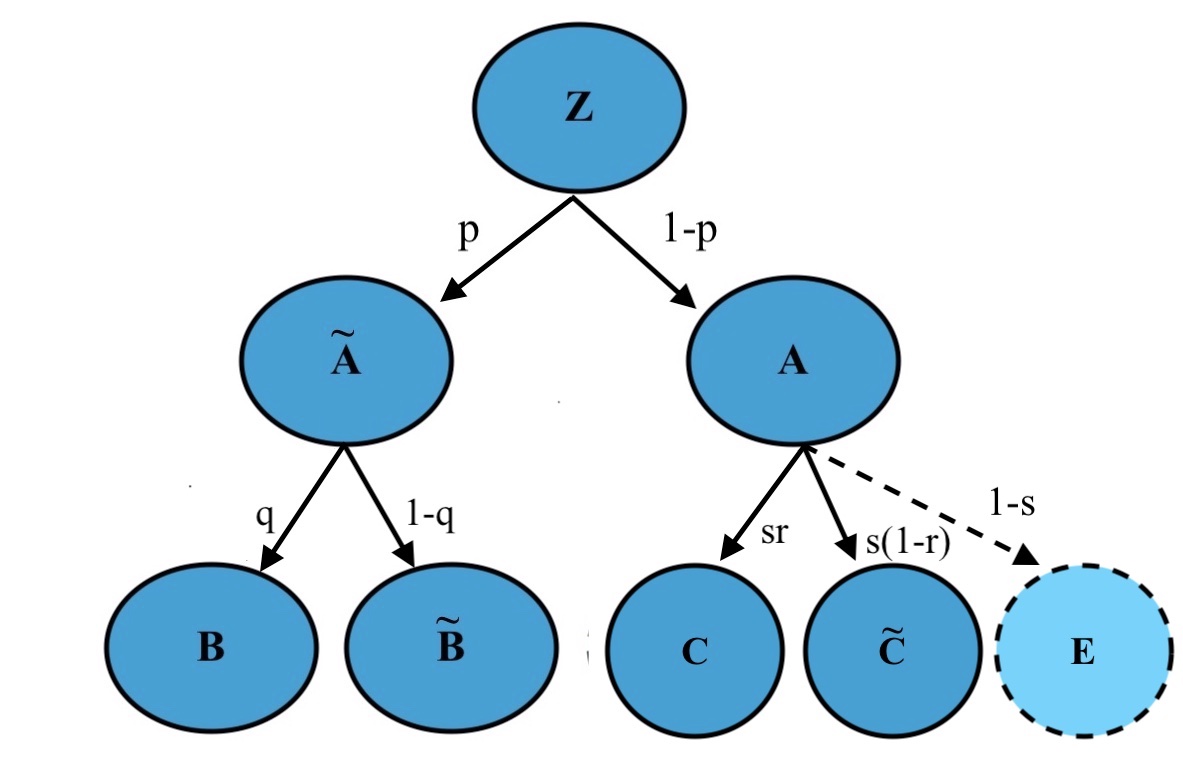}
	\caption[Simple tree used with hidden node.]{Simple tree used with hidden node.  This tree is an extension of Figure \ref{fig:simstree} which assumes that data is available for nodes $B$, $C$, $\tilde{C}$ as opposed to $A, B$.  Node $E$ is latent, and represents the error known to exist in observed values of $C$ and $\tilde{C}$.  An estimate of population proportion missed from $A$ is assumed known, so that  probabilities $r, 1-r$ are proportionally reduced.}
	\label{fig:simstree1}
\end{figure}

Now suppose that instead of having data for node $A$, marginal counts exist for two descendent nodes of $A$, $C$ and $\tilde{C}$, resulting in the model in Figure \ref{fig:simstree1}.  If independent branch estimates are available for $p,q,r$ and counts of nodes $C$, $\tilde{C}$ are known with $A = C + \tilde{C}$, then 
\begin{equation}\label{eq:sumnodes}
f(z|b, c, \tilde{c}) = f(z|a,b,c,\tilde{c}) = f(z|a,b),
\end{equation}
since $Z$ is conditionally independent of $C$ and $\tilde{C}$ given node $A$.  Furthermore, by referring to Figure \ref{fig:generalDAG}, we observe that conditioning on node $A$ blocks all dependency paths between nodes $C$, $\tilde{C}$ and $Z$ \cite{murphy3}; sub-trees rooted at an observed node may be dropped from the Bayesian model, so long as estimates for the parameters contained in the sub-tree are not required.  This result further suggests that node counts can be combined among sibling groups in the Bayesian model.  So long as all branches are informed by the same prior data, the same is true for the WMM.  

Due to errors in count data, sibling counts are unlikely to sum to the true parent value; this is the norm in applications involving hard-to-reach target populations in public health and epidemiology which involve health administrative data, where biases and missed counts are well-known.  In this case, aggregating descendent node counts and modeling these as parent values may bias posterior estimates of $Z$, since the posterior distributions of the root population size are not equal in these two frameworks. 
To see this, we use the tree in Figure \ref{fig:simstree1} as a reference for notation and labelling of nodes and branching, without loss of generality.  Let $\tau \sim Beta(\alpha_\tau, \beta_\tau)$, where $\alpha_\tau, \beta_\tau \in \N$ for $\tau \in \{p,q,r,s\}$.
Then the joint posterior is 
\begin{equation}\label{eq:splitjoint}
	f(z,p,q,r,s| b, c, \tilde{c}) \propto f(z)f(p)f(q)f(r)f(s)f(b, c, \tilde{c} | z, p,q,r,s),
\end{equation}
where 
\begin{align*}
	f(b, c, \tilde{c} | z, p,q,r,s) \sim &
	Multi(z, [(pq)^b, ((1-s)r(1-p))^c, ((1-p)(1-r)(1-s))^{\tilde{c}}, 
	(p(1-q)+s(1-p))^{z-b-c-\tilde{c}}])
\end{align*}
Integrating \eqref{eq:splitjoint} with respect to $p,q,r,s$, the integral with respect to $r$ can be simplified, so that for this integral, we have
\[
\int \frac{r^{\alpha_r + c - 1}(1-r)^{\tilde{\beta_r + c -1}}}{B(\alpha_r, \beta_r)} dr.
\]
As in the derivation of \eqref{eq:simplified}, we define $S_r$ such that
\begin{align*}
	S_r &= \frac{\alpha_r \cdot ... \cdot (\alpha_r + c - 1)\cdot \beta_r \cdot ... \cdot (\beta_r + \tilde{c} - 1)}{(\alpha_r + \beta_r) \cdot ... \cdot (\alpha_r + \beta_r +c+\tilde{c}-1)}.
\end{align*}
Since $S_r$ is not a function of $r$ itself, the integral simplifies and we may write the posterior distribution as
\begin{align}\label{eq:hiddennodepost}
	S_r\cdot \frac{f(z)z!}{b!c!\tilde{c}!(z-b-c-\tilde{c})!}\cdot \int \int \frac{p^{\alpha_p + b - 1}(1-p)^{\beta_p + c + \tilde{c}-1}}{B(\alpha_p, \beta_p)}&\cdot \frac{q^{\alpha_q + b - 1}(1-q)^{\beta_q-1}}{B(\alpha_q, \beta_q)}\cdot  
	 \frac{s^{\alpha_s - 1}(1-s)^{\beta_s + c + \tilde{c}-1}}{B(\alpha_s, \beta_s)}\\ \nonumber
  &\cdot(p(1-q)+s(1-p))^{z-b-c-\tilde{c}} dp \text{ } dq \text{ } ds.
\end{align}
When $s=1$, similarly defining $S_p, S_q$ and separating the integral is possible, resulting in a posterior given by
\[
S_pS_qS_r\cdot \frac{f(z)z!}{b!c!\tilde{c}!(z-b-c-\tilde{c})!}.
\] 
Further substituting $c + \tilde{c} = a$, $S_p$ and $S_q$ become equal to $C_p$ and $C_q$, respectively, as in equations \eqref{eq:cp} and \eqref{eq:cq}, so that we have
\begin{equation}\label{eq:sequals0post}
	C_pC_qS_r\cdot \frac{f(z)z!}{b!c!\tilde{c}!(z-a-b)!}.
\end{equation}
Then since $\alpha_r,\beta_r$ are independent of $z$,
\begin{align*}
	\frac{S_r}{c!\tilde{c}!} 
	&\propto \frac{1}{a!},
\end{align*}
equation \eqref{eq:sequals0post} is proportional to equation \eqref{eq:simplified}, as we expect from equation \ref{eq:sumnodes}.
However, for $s\neq 1$, a closed-form of the integral equations \eqref{eq:hiddennodepost} is not readily available.  
This result suggests that when marginal counts are uncertain, incorporating a hidden node is a more representative a structure, and a model should incorporate these missed counts, given the posterior distributions of equations \eqref{eq:hiddennodepost} and \eqref{eq:simplified} are not equal; in addition, a hidden node may also be considered a possible \textit{cause} of uncertain marginal counts.  Inclusion of latent nodes representing these known sources of error may provide improvements to root node size estimation, advantaging a Bayesian model over the WMM when researchers are aware of count uncertainty.

\subsubsection{Methods Comparison and Validation}
\begin{figure}
	\centering
	\includegraphics[height=0.4\textheight]{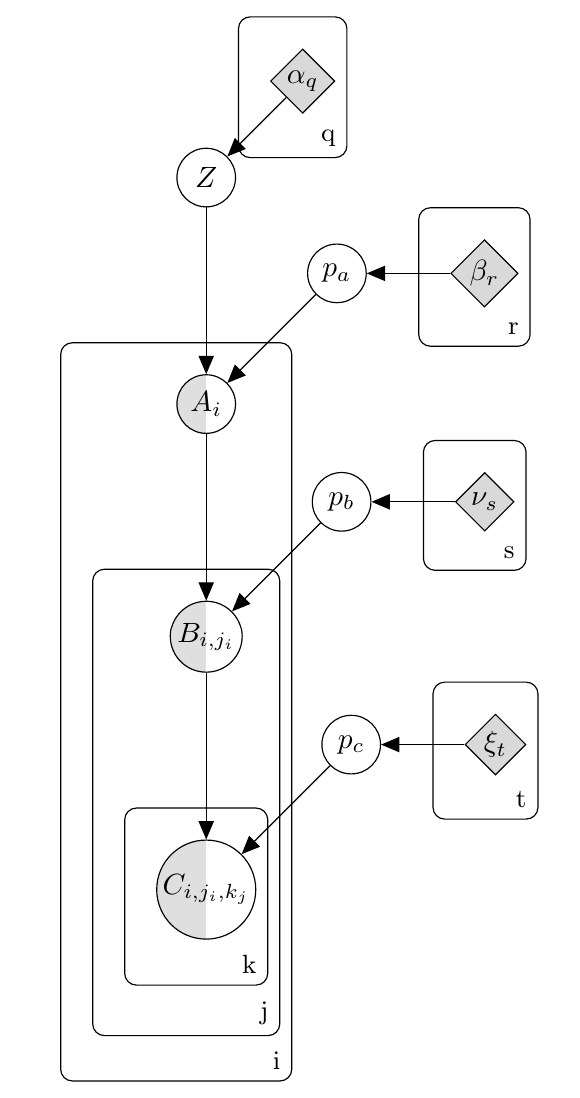}
	\caption[Graphical model for general tree-structured data with three levels. ]{Graphical model for general tree-structured data with three levels.  Shaded nodes represent parameters derived from prior knowledge or literature, with half-shaded nodes representing that data may or may not be available at these nodes, depending on the indices.  Indices $i,j_i,k_j \in \N$, where $j,k$ differ depending on the number of child nodes of each parent.  The dimensions of hyperparameters $q,r,s,t \in \N$ differs also based on the chosen distributions, but we assume these are known hyperparameter inputs to the priors of $Z$, $p_a$, $p_b$, and $p_c$.}
	\label{fig:generalDAG}
\end{figure}

Processing survey data can be done as part of one complete generative model or as a two-step procedure, the first step involving the determination of posterior branching distributions given uninformative priors and survey data.  
Referring to Figure \ref{fig:generalDAG} as a representative model, the hyperparameters $\{\alpha_q\}$, $\{\beta_r\}$, $\{\nu_s\}$, $\{\xi_t\}$, may be informed by using survey values as directly, or chosen subjectively.  Let $D_\mathcal{L^*}$ denote the marginal count data of observed leaves, and let $D_\mathcal{S}$ denote the survey data used to inform the sets of  hyperparameters.  Then a generative model for the data and parameters under the assumption that branching survey data is independent of marginal count data is
\begin{align}\label{eq:jointprocessing1}
f(z, p_a, p_b, p_c, D_\mathcal{L^*}, D_\mathcal{S}) = &f(D_\mathcal{L^*} | z, p_a, p_b, p_c) \cdot f(D_\mathcal{S}| p_a, p_b, p_c)\cdot f(z) \cdot f(p_a, p_b, p_c),
\end{align}
where we also apply independence of $Z$ and branching distributions.  Thus the joint conditional distribution satisfies 
\begin{align}
f(z, p_a, p_b, p_c | D_\mathcal{L^*}, D_\mathcal{S}) \propto &f(D_\mathcal{L^*} | z, p_a, p_b, p_c) \cdot f(D_\mathcal{S}| p_a, p_b, p_c) \cdot f(z) \cdot f(p_a, p_b, p_c)\label{eq:jointprocessing2} \\
&\propto f(D_\mathcal{L^*} | z, p_a, p_b, p_c) \cdot f(p_a, p_b, p_c | D_\mathcal{S}) \cdot f(z) \label{eq:jointprocessing3}.
\end{align}
The above simultaneously processes survey data used to inform the priors of branching distributions.  While equation \eqref{eq:jointprocessing3} gives the simplest form for MCMC sampling, there are some situations for which equations \eqref{eq:jointprocessing1} and \eqref{eq:jointprocessing2} must be used instead. 

The WMM and a Bayesian model on an equivalent data structure have a number of comparable attributes; both methods use prior knowledge to inform branching probabilities, and though the evolution of sampling differs, some specifications of the branching priors facilitate more direct comparison between the two methods.  In particular, consider the first step of the two-step Bayesian model described above, in which branching data is pre-processed to generate posterior distributions, which are then used as priors in the second stage of the Bayesian model:
\[
f(br | D_{\mathcal{S}}) \propto f(br) \cdot f(D_{\mathcal{S}} | br)
\]
For a uniform $br \sim Beta(1,1)$ prior on branching with $D_{\mathcal{S}} | br \sim Bin(N, br)$, the above posterior is
\[
br | D_{\mathcal{S}} \sim Beta(x + 1, N-x+1),
\] 
which is the resulting distribution when survey values as used directly as branching hyperparameters.  The conjugacy of the $Dirichlet$ prior to the $Multinomial$ distribution gives a similar result for non-binary branching.  In particular, marginal branch probabilities are given by 
\[
p_{N_i} \sim Beta\left(N_i + 1, \sum_{j \in \mathcal{G}} N_j - N_i + 1\right),
\]
so that the WMM and Bayesian methodologies are both ultimately sampling the same posteriors $f(br | D_{\mathcal{S}})$ after the first step of the above procedure; the second step of the Bayesian model, however, involves a root node prior and the likelihood of the count data in place of a fixed value, and it is here that the methods differ.  
In addition, where the set of latent nodes or branches is non-empty, the Bayesian model can still incorporate what data are available, while the WMM cannot utilize data which are not part of an informative path.  Furthermore, while the implementation of the WMM synthesizes the evidence from multiple surveys informing a sibling group of branches by adjusting the joint distribution with rejection schemes and importance sampling \cite{flynncomp}, a number of techniques to choose an appropriate $Dirichlet$ distribution for the sibling group in the Bayesian model could be employed.

\subsection{Simulation}
A set of initial conditions and specifications for the model is obtained by processing observations and choosing hyperparameters, after which MCMC methods can be used to obtain posterior distributions on parameters from the model for estimation purposes.  Analysis has been conducted in JAGS \cite{jags}, a Gibbs sampler using MCMC simulation designed to analyze hierarchical Bayesian models.  Where Gibbs sampling strategy is not feasible, however, JAGS automatically resorts to other sampling techniques \cite{jags}.  A variety of software and probabilistic programming languages exist which are similarly suitable to solving hierarchical Bayesian models with MCMC, such as Stan \cite{stan} and Blang \cite{blang}.  Methods for providing analytic approximations to the posterior, such as variational Bayesian methods, can also be explored as alternatives to the above, and may provide faster solutions with comparable accuracy \cite{irvinevarbayes2019}, especially with increasing model complexity.

The simulated experiment uses the tree structure in Figure \ref{fig:simstree}.  We let $Z=1000$ be the true root node population size, with prior $f(z)$ assuming either a uniform prior with limits $u, v$, or a normal distribution with mean $\mu$ and variance $\sigma^2$ in each trial $i \in \{1,...,r\}$, and the number of trials $r=10000$ constant across experiments.  We choose $p=0.25$, $q=0.8$; together these parameters induce observed values of $A,B$.  We also adjust the survey size, $S$.  This value is assumed constant and represents the sample size of each simulated survey used to inform $p,q$, and is adjusted across experiments to explore effects of evidence quality.  We then generate $n$ fictitious random survey results per trial, $s_{p,i}, s_{q,i}$, for branches $p,q$, respectively, by sampling  $s_{p,i} \sim Binom(S, p)$, $s_{q,i} \sim Binom(S, q)$ for each $i \in [1,...,n]$, obtaining
\begin{align}
	\hat{p}_i &= \frac{\hat{s}_{p,i}}{S}, \text{   }\hat{q}_i = \frac{\hat{s}_{q,i}}{S}.
\end{align}
Samples $(\hat{A}_i, \hat{B}_i)$ can be generated using a $Multi(1-p, pq, p(1-q))$ distribution, or in a two-step procedure by setting $A_i \sim Bin(Z, 1-p)$, $B_i \sim Bin(Z-\hat{A}_i, q)$.
We have two sources of trial-to-trial variation - the survey values informing the distributions of $p, q$, as well as the values of leaves $A,B$.  The WMM generates multiple samples of $p,q$ from $Beta(\alpha,\beta)$ distributions to construct a covariance matrix and calculate optimal weights using log-transformed data, and thus a point estimate $\widehat{\log(Z)}_i$ for each trial $i$.  The Beta parameters are assumed to be informed by survey or literature estimates of each branch, so we set $\alpha_{p,i} = s_{p,i}+1$, $\beta_{p,i} = S-s_{p,i}+1$, and similarly for $f(q)$.  These parameters are similarly used for each trial of the MCMC modeling in JAGS, as well as in the closed-form distribution.  For each of these methods, the prior density on $Z$ varies by experiment, and we try both a uniform density as well as a Gaussian density. Root-mean-squared errors (RMSE) are calculated in each case by summing the squared difference of $\log Z$ and estimates $\widehat{\log(Z)}_i$ for each trial, $i$.  
The full details of implementation can be found in the supplementary material. 

Sampling branches independently from leaf-to-root has computational advantages; samples may be drawn at the time a branch is reached in back-calculation, overwriting the previous value and reducing required memory, since sibling branches summation constraints are ignored.  
Over many iterations, the effect of ignoring this constraint may not significantly affect the outcome; however, this method is likely more sensitive to bias incurred by inaccurate prior knowledge on a subset of paths, which may be given unjustly high weight if priors are moderately precise, while priors could be incongruent with other sibling estimates. 
By contrast, incorporating importance sampling and rejection schemes satisfies the constraints on sibling branch groups, providing a more accurate model of the tree structure.  These schemes could result in prohibitively long compute times, but they may also produce estimates more consistent with Bayesian modeling.  
We compare both the independent $Beta$ sampling of paths and the mixed scheme of branch sampling, which uses $Dirichlet$ distributions where possible and importance sampling or rejection schemes where necessary. The two methods are referred to in what follows as \textit{WMM-Ind} and \textit{WMM-Dir}, respectively.

We wish to compare the estimate given using the WMM to those obtained using a Bayesian model.  To account for any simulation error with MCMC, we also include sampling directly from the closed-form posterior distribution.  We expect correlation with Bayesian estimates to be greater with the WMM-Dir sampling scheme than with the WMM-Ind sampling.  We also wish to examine whether there exists a setting in which the WMM may outperform MCMC.  Since MCMC results represent the closed-form solution with some numerical approximation error, we hypothesize that MCMC will correlate more closely with the closed-form solution where the true value of $Z$ is consistent with the prior distribution, but that WMM may gain an advantage with a poor choice of prior on $Z$.   In general, we expect the both WMM-Ind and WMM-Dir to also perform better with higher sample size, $S$, as variability of root population size estimates is tied to this value through the $Beta/Dirichlet$ branching parameters.  

\subsection{Estimating HCV Prevalence}
One population which remains partially hidden from healthcare administrative data sources is the population of individuals infected with Hepatitis C virus (HCV), though the virus is a leading cause of chronic liver disease \cite{dore2002}.  Methods of determining the size of the underlying infected population often rely on CRC or multiplier methods approaches, and data on number of infections and number of injection drug users are often used to make inference on the other hidden population \cite{prevostHCV2015,sweetingIDU2009,jonesCRC2014,degenhardt2016}. 

\begin{figure}
	\centering
	\includegraphics[width=\linewidth, height=0.27\textheight]{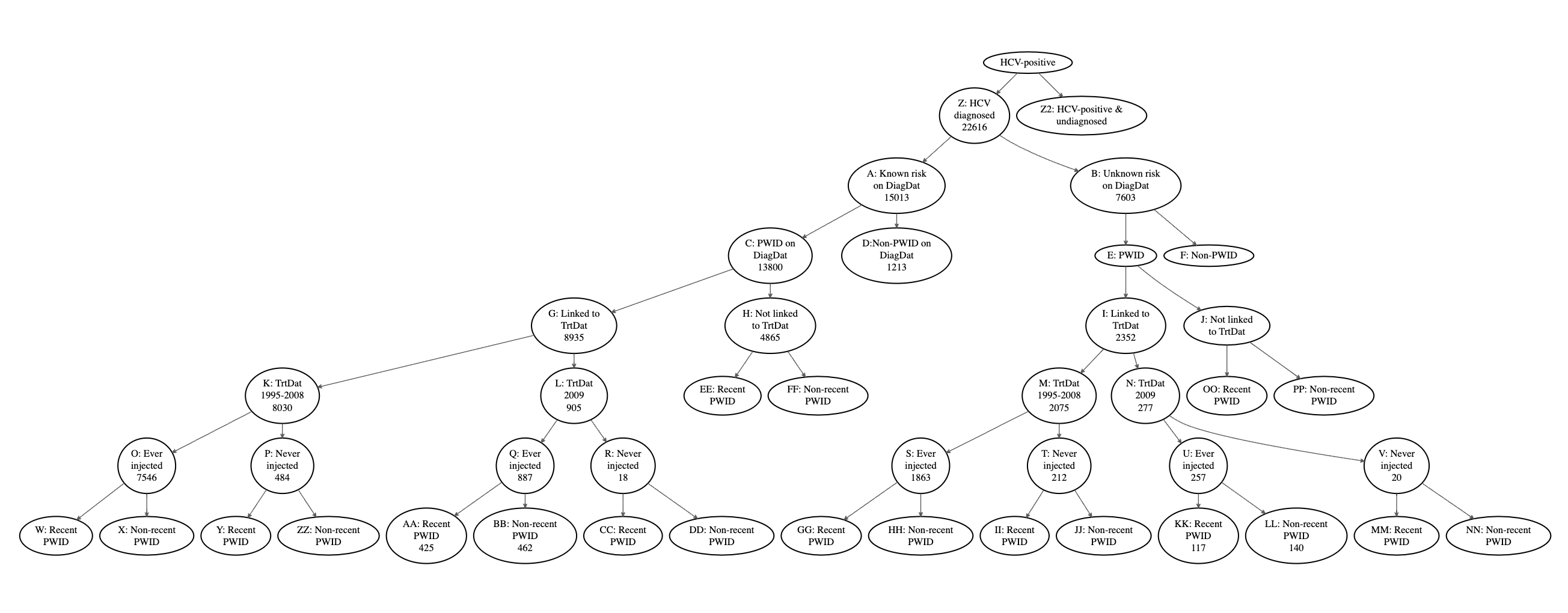}
	\caption{Tree-structure representing available data to estimate the number of HCV-infected persons in Scotland in 2009 \cite{prevostHCV2015}. Counts stated in node where available.}
	\label{fig:deangelistree}
\end{figure}

Prevost et al. \cite{prevostHCV2015} applied a Bayesian approach to estimate the number of people who inject drugs (PWID) and are HCV-prevalent in Scotland in 2009, and the infected PWID population who remains undiagnosed.  The relationships between sources of available data present as tree-structured, and estimates are available to extent this tree to estimate all HCV positive individuals (both PWID and non-PWID) by projecting back a further ancestral generation to a root node which is defined by this population (see Figure \ref{fig:deangelistree}).  We then test an application of the WMM on the data available to inform Figure \ref{fig:deangelistree}.
A summary of parameter values used for each of the nodes can be found in Table \ref{table:deangelisparams}.  Though prior distributions were chosen for branching were already chosen for the Bayesian approach in Prevost et al. \cite{prevostHCV2015}, some choices differ herein; in particular, we parameterize branching $Beta$ distributions using administration data counts where available for the WMM, which differ from the uniform priors place on several of these branches in the Bayesian model \cite{prevostHCV2015}. Since the WMM approach does not update the prior via a likelihood to generate a posterior distribution, we include the available data in parameterization of the distribution itself.  In addition, $\alpha_Z, \beta_Z$ were chosen to reflect that this value is estimated to be approximately 42\%, with bounds between 30\% and 57\%.  Similarly, $\alpha_E, \beta_E$ were chosen to reflect a $Beta$ distribution with mean approximately at 0.65, the estimated value of this branching probability.  Since no data or past literature estimates were provided for $p_I$, this was set to be uniformly distributed; model estimates and variability are likely to be improved by refining this choice of distribution.

\section{Results}
\subsection{Simulated Experiment}
A summary of parameter values explored in a variety of numbered experiments can be found in Table \ref{table:simparams}; experiments are referred to by number in what follows.
\begin{table}
	\centering
	\begin{tabular}{|c|c|c|c|c|}
			\hline
			Experiment & Prior & $\gamma$ & $\eta$ & $S$ \\
			\hline \hline
			1 &  $unif$ & 750 & 1250 & 50 \\
			2 & $unif$ & 0 & 10000 & 50\\
			3 &  $unif$ & 750 & 1250 & 1000 \\
			4 & $unif$ & 0 & 10000 & 1000 \\
			5 & $Gauss$ & 2000 & 150 & 50 \\
			\hline
		\end{tabular}
	\caption[Summary of parameters used in each simulation experiment.]{Summary of parameter values used in each simulation experiment.  $\gamma$, $\eta$ represent the parameters of the prior $f(z)$, so that when $f(z)$ is uniform, these correspond to the endpoints of the uniform bounds, $\gamma = u$, $\eta = v$, and when $f(z)$ is Gaussian, $\gamma = \mu$ and $\eta = \sigma$.  $S$ corresponds to survey sample size.  In all experiments, $Z=1000$, $p=0.25$, $q=0.8$, and $r=10000$.}
	\label{table:simparams}
\end{table}
The first experiment uses $S = 50$ and a prior $Z \sim Unif(u,v)$ for the closed-form analysis and MCMC, with $u = 750$, $v=1250$.  The resulting MSE and RMSE for each method under these settings can be found in Table \ref{table:MSEexp1}.  Correlation plots comparing results of the four methods using log-transformed data can be seen in Figure \ref{fig:corplotsexp1} of the supplementary material.

\begin{table}
	\centering
	\begin{tabular}{|c|c|c|c|c|c|}
		\hline
		Method & \multicolumn{5}{|c|}{Experiment} \\ 
        \cline{2-6}
         & 1 & 2 & 3 & 4 & 5 \\
		\hline \hline
		Closed-form & 2.06e-2 & 2.07e-2 & 8.37e-3 & 8.48e-3 & 2.06e-2 \\
		MCMC & 2.21e-2  & 2.23e-2 & 8.86e-3 & 8.96e-3 & 6.30e-2 \\
		WMM-Ind & 4.87e-2 & 4.82e-2 & 1.63e-2 & 1.66e-2 & 4.82e-2 \\
		WMM-Dir & 2.41e-2 & 2.45e-2 & 8.73e-3 & 8.67e-3 & 2.40e-2 \\
		\hline
	\end{tabular}
	\caption{RMSE errors associated with each of the five experimental conditions in Table \ref{table:simparams} and the four methods to estimate $\log Z$.}
	\label{table:MSEexp1}
\end{table}

\begin{figure}
	\centering
	\includegraphics[height=0.35\textheight]{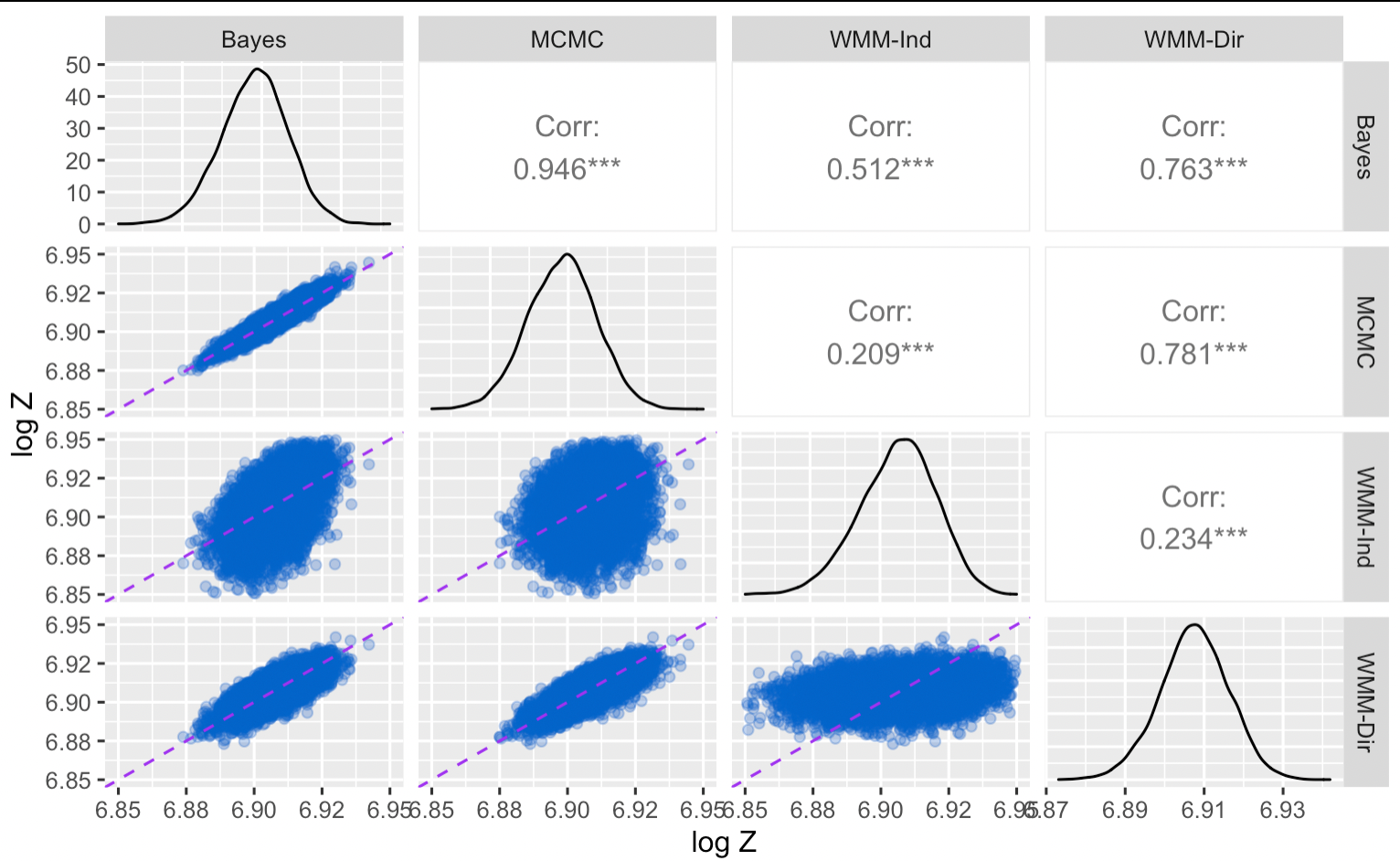}
	\caption[Simulation experiment 3 correlation plots comparing $\log Z$ estimates generated by closed-form Bayes solution, MCMC, and both methods of WMM.]{Experiment 3 - Correlation plots comparing $\log Z$ estimates generated by the closed-form Bayes solution, MCMC, and WMM methods in experiment 3, where prior $Z \sim Unif(750, 1250)$ and survey size $S=1000$.}
	\label{fig:corplotsexp5}
\end{figure}

Additional scenarios were explored to assess the effects of prior choice $f(z)$, as well as a proxy for assessing the effect of evidence quality by way of adjusting the survey size $S$, where an increase in $S$ decreases variation in branch samples.  We first adjust the prior in experiment 2 so that $u=0$, $v=10000$, keeping $S=50$.  These results can be found in Table \ref{table:MSEexp1}, with correlation plots in Figure \ref{fig:corplotsexp4} of the supplementary materials.  The third experiment uses the original $Unif(750,1250)$ prior but adjusts sample size to $S=1000$, resulting in the RMSE values in Table \ref{table:MSEexp1}, and the correlation plots of Figure \ref{fig:corplotsexp5}, presented herein as an example correlation plot.  To investigate the scenario where we have stronger evidence on branching but a weaker prior on $Z$, we set $S=1000$, $u=0$, and $v=10000$ in experiment 4; the resulting RMSE of each method can be found in Table \ref{table:MSEexp1}, with correlation plots for log-transformed estimates in Figure \ref{fig:corplotsexp2} of the supplementary.

The last experiment explores the comparison across methods in the event that a more decisive, but inaccurate prior is chosen.  To illustrate this scenario, we choose the prior $f(z)$ to be Gaussian with $\mu = 2000$ and $\sigma = 150$.  We also set the survey size back to $S=50$.  
This distribution may be chosen if, for instance, we believe $Z \in (1800,2200)$ and have roughly 80\% confidence in this interval, so that the true value is sampled with low probability.  The results in this setting can be found in Table \ref{table:MSEexp1}, with correlation plots in the log-transformed and untransformed case in Figure \ref{fig:corplotsexp3} of the supplementary material.  Density plots and trace plots of $N$, $p$, $q$ can be found in Figures \ref{fig:wrongpriortrace} and \ref{fig:wrongpriordens} in the supplementary materials.

\subsection{Estimating HCV Prevalence}
Under this model, the WMM combines evidence from all five leaves with known counts in Figure \ref{fig:deangelistree} to estimate the total number of HCV-positive individuals in Scotland in the given time period to be 56813, with the central 95\% of the density of samples between 41788 and 82384.  Weights for each path can be found in Table \ref{table:exweights}, with the estimate dominated by the evidence provided by path ending in node $D$.  This is unsurprising, as shorter path lengths incorporate fewer sampled branches, and thus are inherently less variable under otherwise equal branching distributions. 
\begin{table}
	\centering
	\begin{tabular}{|c|c|c|}
        \hline
        Branching probability $p_i$ & $\alpha_i$ & $\beta_i$ \\
        \hline \hline
        $p_Z$ & 20 & 30 \\
        $p_A$ & 15013 & 7603 \\
        $p_C$ & 13800 & 1213 \\
        $p_E$ & 3 & 1 \\
        $p_G$ & 8935 & 4865 \\
        $p_I$ & 1 & 1 \\
        $p_K$ & 8030 & 905 \\
        $p_M$ & 2075 & 277 \\
        $p_O$ & 7546 & 484 \\
        $p_Q$ & 887 & 18 \\
        $p_S$ & 1863 & 212 \\
        $p_U$ & 257 & 20 \\
        $p_{AA}$ & 425 & 462 \\
        $p_{KK}$ & 117 & 140 \\
        \hline
	\end{tabular}
	\caption{Summary of distribution parameter values used to apply the WMM to estimate HCV prevalence in Scotland in 2009 \cite{prevostHCV2015}. Each branching probability leading to node $i$ is $p_i \sim Beta(\alpha_i + 1, \beta_i + 1)$.}
	\label{table:deangelisparams}
\end{table}
\begin{table}
	\centering
	\begin{tabular}{|c|c|}
		\hline
		Path Endpoint & WMM Weight  \\
		\hline \hline
		$AA$ & 0.235 \\
		$BB$ & 0.176 \\
		$D$ & 0.579 \\
		$KK$ & 0.017 \\
		$LL$ & -0.008 \\
		\hline
	\end{tabular}
	\caption{Variance-induced weights generated by the WMM for all paths ending in leaf counts in the tree in Figure \ref{fig:deangelistree}.}
	\label{table:exweights}
\end{table}

\section{Discussion}
For experiment 1, which used $Z \sim Unif(750, 1250)$ and $S=50$, the RMSE was approximately 2.3 times larger for WMM-Ind as compared to the closed-form Bayes, and 2.2 times larger than the MCMC's RMSE.  
A similar discrepancy in RMSE was seen even with even weaker $Uniform$ bounds in experiment 4, suggesting decreased prior confidence was not sufficiently advantageous to put WMM-Ind on par with MCMC.  This result is perhaps not surprising - the WMM does not use likelihood methods, which are maximally efficient. 
Improved RMSE were universally observed using the WMM-Dir method; in all experiments, RMSE were on the same order of magnitude as the sampling from the closed-form solution.

As expected, a strong correlation is observed between MCMC and the closed-form Bayes solution; differences are consistent with expected errors due to Monte Carlo sampling and rounding, as the discrete uniform distributions have been approximated using continuous uniform distributions in JAGS.  Positive correlation between WMM-Ind and the closed-form Bayes solution is moderate in some settings, as in Figure \ref{fig:corplotsexp5}.  Lesser positive correlation is observed between MCMC and WMM-Ind.  Though increasing the sample size from $S=50$ to $S=1000$ did not improve the overall RMSE of the WMM-Ind relative to MCMC, it did improve the correlation between the WMM-Ind and the closed-form Bayesian solution, and the distributional agreement between the estimates of WMM-Ind and the closed-form Bayes solutions is visually improved in these setting (see Figures \ref{fig:corplotsexp5} and \ref{fig:corplotsexp2} of the supplementary).  

\begin{figure}
	\centering
	\includegraphics[height=0.35\textheight]{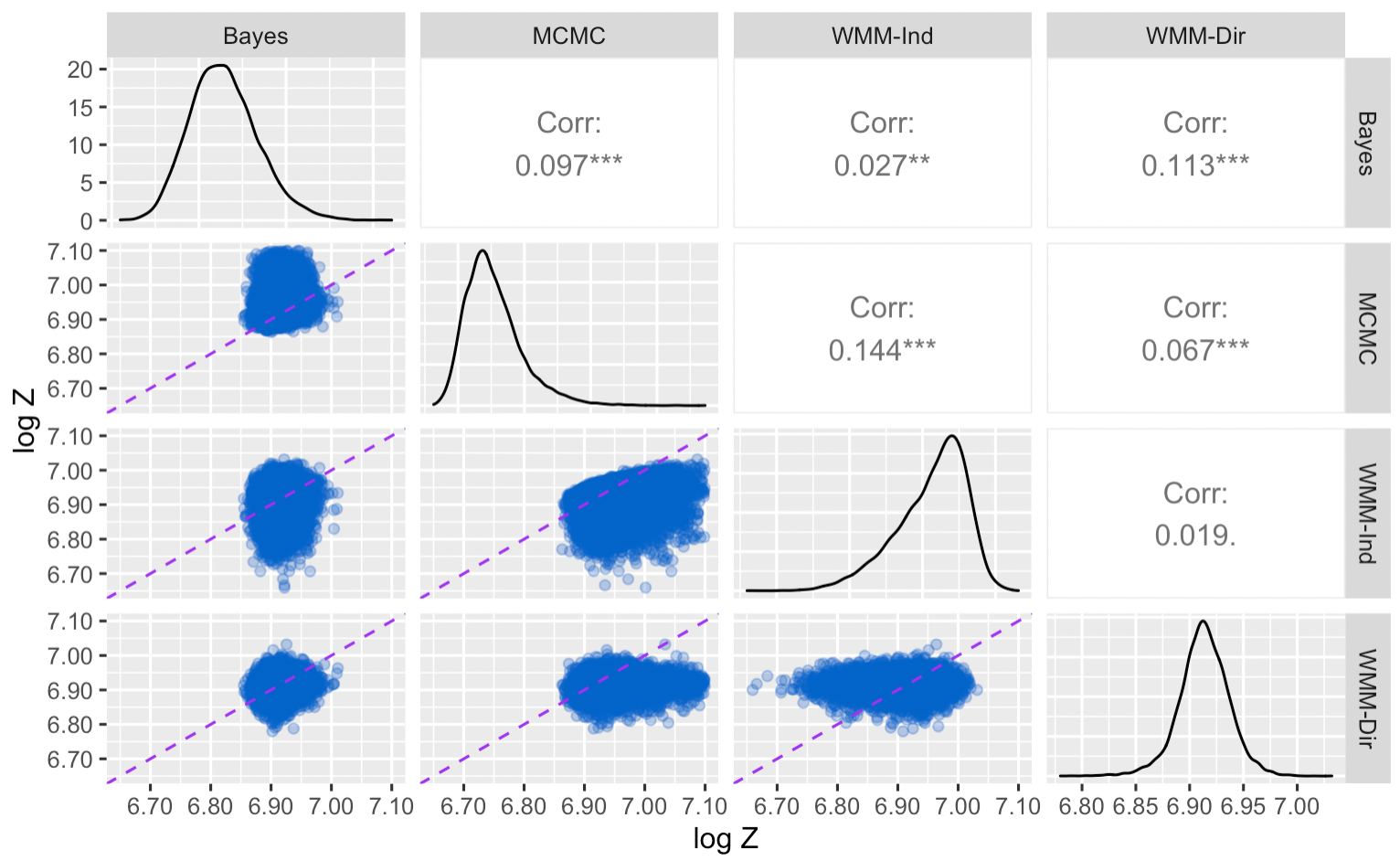}
	\caption[Simulation experiment 5 correlation plots comparing $\log Z$ estimates generated by closed-form Bayes solution, MCMC, and both methods of WMM.]{Experiment 5 - Correlation plots comparing $\log Z$ estimates generated by the closed-form Bayes solution, MCMC, and WMM methods in experiment 5, where prior $Z \sim N(2000, 150)$ and survey size $S=50$.}
	\label{fig:corplotsexp3}
\end{figure}

\begin{figure}
	\centering
	\includegraphics[width=\linewidth,height=0.4\textheight]{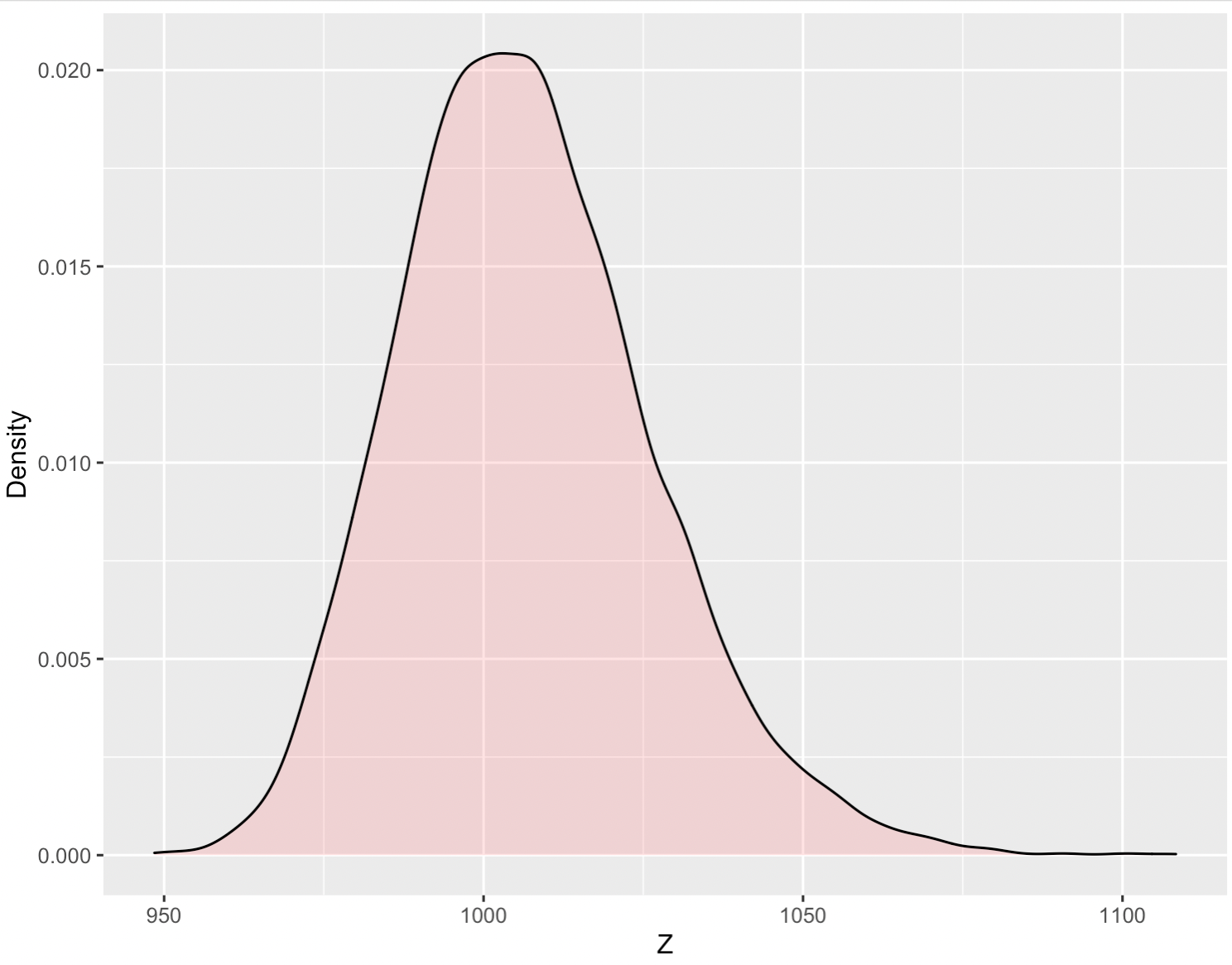}
	\caption{Experiment 5 - Density plot of $Z$ from the closed-form Bayesian solution.}
	\label{fig:wrongpriorcfdens}
\end{figure}

Experiment 5 explored a scenario which disadvantages MCMC, while the WMM remains unaffected since the model does not incorporate a root node prior.  We expected this to impact the quality of MCMC estimates, possibly distinguishing a scenario where the results of WMM may be preferable to MCMC.  This prior does appear to sufficiently hinder the performance of MCMC; we see a significant effect on the RMSE of MCMC, and the WMM outperforms MCMC in this case, with the RMSE of the WMM-Dir method showing an order of magnitude improvement over MCMC.  Trace and density plots from one of the 10,000 iterations of the MCMC from experiment 5 and the density plot from the closed-form can be found in the supplementary (Figures \ref{fig:wrongpriortrace}, \ref{fig:wrongpriordens}, \ref{fig:wrongpriorcfdens}, respectively).  Effective sample sizes of $Z$, $q$, and $p$ were 6000, 6000, and 3000, respectively.  The autocorrelation function (ACF) plot, can also be found in Figure \ref{fig:acfsim} of the supplementary.  Despite the poor statistical properties of the posterior distribution resulting from the choice of prior, these values and plots seem to suggest the posterior has been well computed and the MCMC has converged relatively well among the six chains, though some evidence of bimodality in the posterior distribution of $Z$ can also be seen (see Figure \ref{fig:wrongpriordens} in supplementary).  The posterior distribution of $Z$ is heavy-tailed, which may help to explain the higher estimates among the MCMC in comparison to the closed-form Bayes solution.  Differences between these outcomes may also be partly explained by the rounding procedure which converts the continuous $Normal$ prior on $Z$ to a discrete distribution, but some error does appear to be unexplained by these diagnostics.  In addition, the correlations among all methods are lower, including between MCMC and the closed-form solution, as seen in Figure \ref{fig:corplotsexp3}.  
This experimental setting demonstrates that an uninformative prior containing the true value of the root is sufficient for MCMC to outperform WMM; however, a poor guess at the prior which assign low probability to the true root population value may sufficiently bias MCMC, such that the RMSE is worse than that of the WMM.  While this is an interesting result, it is of little practical significance, as a scenario in which we have high confidence in a bad choice of prior would not be detectable in practice.  However, if prior knowledge is possibly biased, these results do support either the use of methods which do not rely on a root prior, such as the WMM, or an uninformative root prior in the Bayesian model.

The above simulations have been performed on a simplified tree, in which a closed-form solution was attainable.  In practice, we are unlikely to have access to a closed-form solution and modeling and approximation methods are required for estimation.  While the RMSE of MCMC is significantly better than that of WMM-Ind even with a weak prior, there may be practical scenarios in which the WMM-Ind method may still be considered for estimation.  For example consider the RMSEs of experiment 4 in Table \ref{table:MSEexp1}.  While an RMSE of $8.48 \times 10^{-3} \approx ln(1.0085)$ suggests the model values will be roughly a factor of $1.0085$ out on average for the MCMC, the RMSE of $1.66 \times 10^{-2}$ suggests a multiplicative factor of only $1.0167$ for WMM-Ind.  Similarly, converting these RMSE values to reflect untransformed population values results in a value of $16.59$ under the WMM-Ind versus $8.96$ for MCMC.  If RMSE is considered heuristically as an estimator of the standard deviation of the error, researchers may determine that a value of $16.59$ for a population size of $Z=1000$ is acceptable, especially if their expertise does not makes the construction of a bespoke Bayesian model for a complex application infeasible.  Furthermore, the computational advantages afforded by independently sampling branch probabilities may be considerable, particularly when trees are large and complex or when sources informing sibling branch data do not agree.  A stronger case can be made for using WMM-Dir as an alternative to Bayesian methods, and these simulations support using the WMM-Dir sampling approach as the default WMM method.  

In the real world example, we compare to the previously published Bayesian model on this data \cite{prevostHCV2015}, which estimates HCV prevalence to be 46657, with 95\% credible interval (33812, 66803).  The estimate of the WMM and associated interval, represented by the central 95\% given by the sample quantiles, could be improved and narrowed, respectively, with further refinement of distributions or counts.  This could be achieved by incorporating past literature estimates or accounting for known biases. Regardless of the simplified approach to this application, the implementation of the WMM was straightforward, with the WMM estimate falling within the credible interval of the Bayesian model. 

Lastly, there are scenarios in which a Bayesian model is likely to apply more readily than the WMM.  For instance, when marginal leaf counts are available but inadequate branching knowledge means these paths are not informative, the WMM cannot be implemented.  Furthermore, when branching knowledge is crude or unrepresentative, the Bayesian model can assign uninformative priors, while the WMM is not well equipped to handle the errors or additional variation associated with uninformative priors.  
A Bayesian model is similarly able to incorporate priors on latent nodes.  This additional flexibility may incur a significant advantage when branching priors are uninformative, as this information could affect the plausible values of branching probabilities along those paths.  
Due to the possibility of incorporating more data and associated variation, the Bayesian model will produce more reliable confidence intervals and will likely achieve more accurate point estimates of root population size in a number of situations.

More experimentation is needed to resolve specifics around the behaviour of the WMM in settings using uninformative priors on either nodes or branches, but extending the methodology to also include distributions on node values is a natural next step that may add considerable value.  Furthermore, while relatively variable paths are down-weighted using the current WMM methodology, another valuable extension may be to further penalize paths whose means estimate for the root population size is significantly different relative to estimates given by other paths.

\section{Conclusion}
The WMM, a method of back-calculation incorporating evidence synthesis on an underlying tree topology has been developed and assessed as an alternative to Bayesian modeling.  The method relies on establishing informative paths, consisting of marginal leaf node counts and estimates for the branching probabilities along the path back to the root node.  Sibling branch groups are sampled jointly according to the distribution admitted by the source data, and path-specific root population size estimates are obtained by back-calculating using marginal leaf node counts and sampled branch probabilities.  Path-specific variances are then used to determine optimal weights and produce a variance-minimized estimate of the root node population size, given the synthesized evidence.  When comparing the WMM's performance to Bayesian methods in a simulated example using RMSE, the WMM using a mixed sampling scheme performed well as compared to MCMC, suggesting it may be a suitable alternative to bespoke Bayesian modeling when the latter is infeasible.  

\section{Supplementary Material}

\begin{figure}
	\centering
	\includegraphics[height=0.35\textheight]{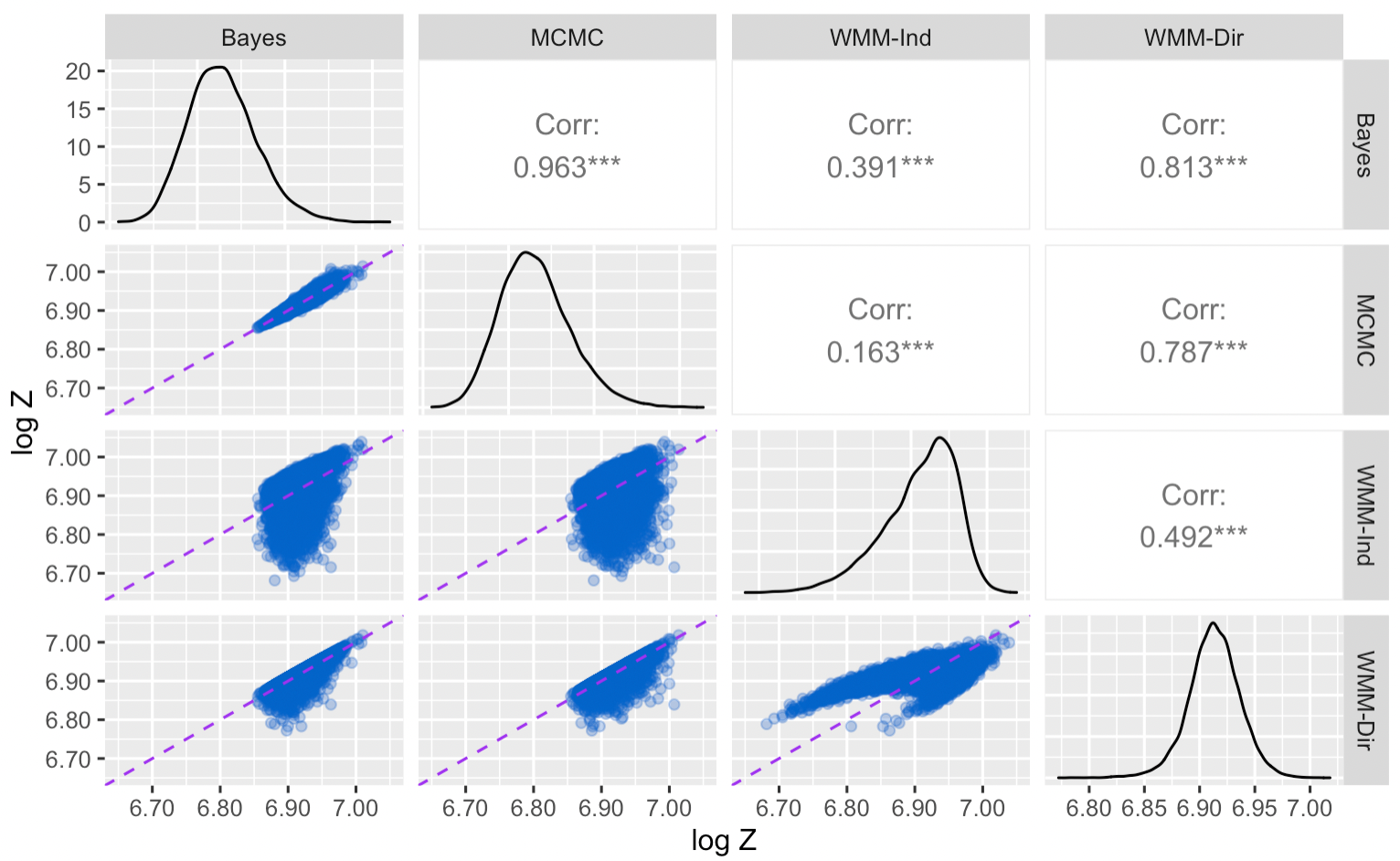}
	\caption[Simulation experiment 1 correlation plots comparing $\log Z$ estimates generated by closed-form Bayes solution, MCMC, and both methods of WMM.]{Experiment 1 - Correlation plots comparing $\log Z$ estimates generated by the closed-form Bayes solution, MCMC, and WMM methods in experiment 1, where prior $Z \sim Unif(750, 1250)$, and survey size $S=50$.}
	\label{fig:corplotsexp1}
\end{figure}

\begin{figure}
	\centering
	\includegraphics[height=0.35\textheight]{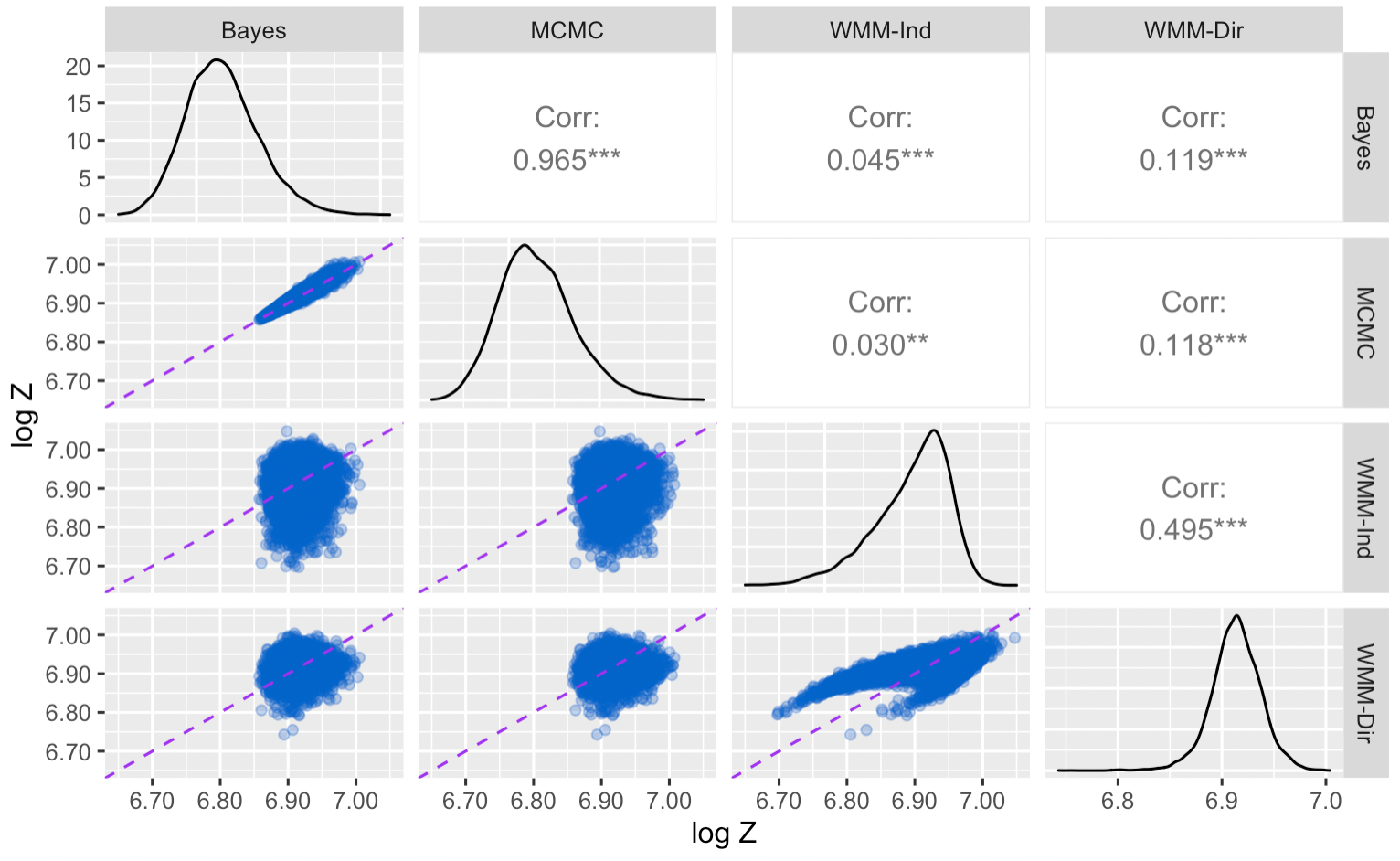}
	\caption[Simulation experiment 2 correlation plots comparing $\log Z$ estimates generated by closed-form Bayes solution, MCMC, and both methods of WMM.]{Experiment 2 - Correlation plots comparing $\log Z$ estimates generated by the closed-form Bayes solution, MCMC, and WMM methods in experiment 2, where prior $Z \sim Unif(0, 10000)$ and survey size $S=50$.}
	\label{fig:corplotsexp4}
\end{figure}

\begin{figure}
	\centering
	\includegraphics[height=0.35\textheight]{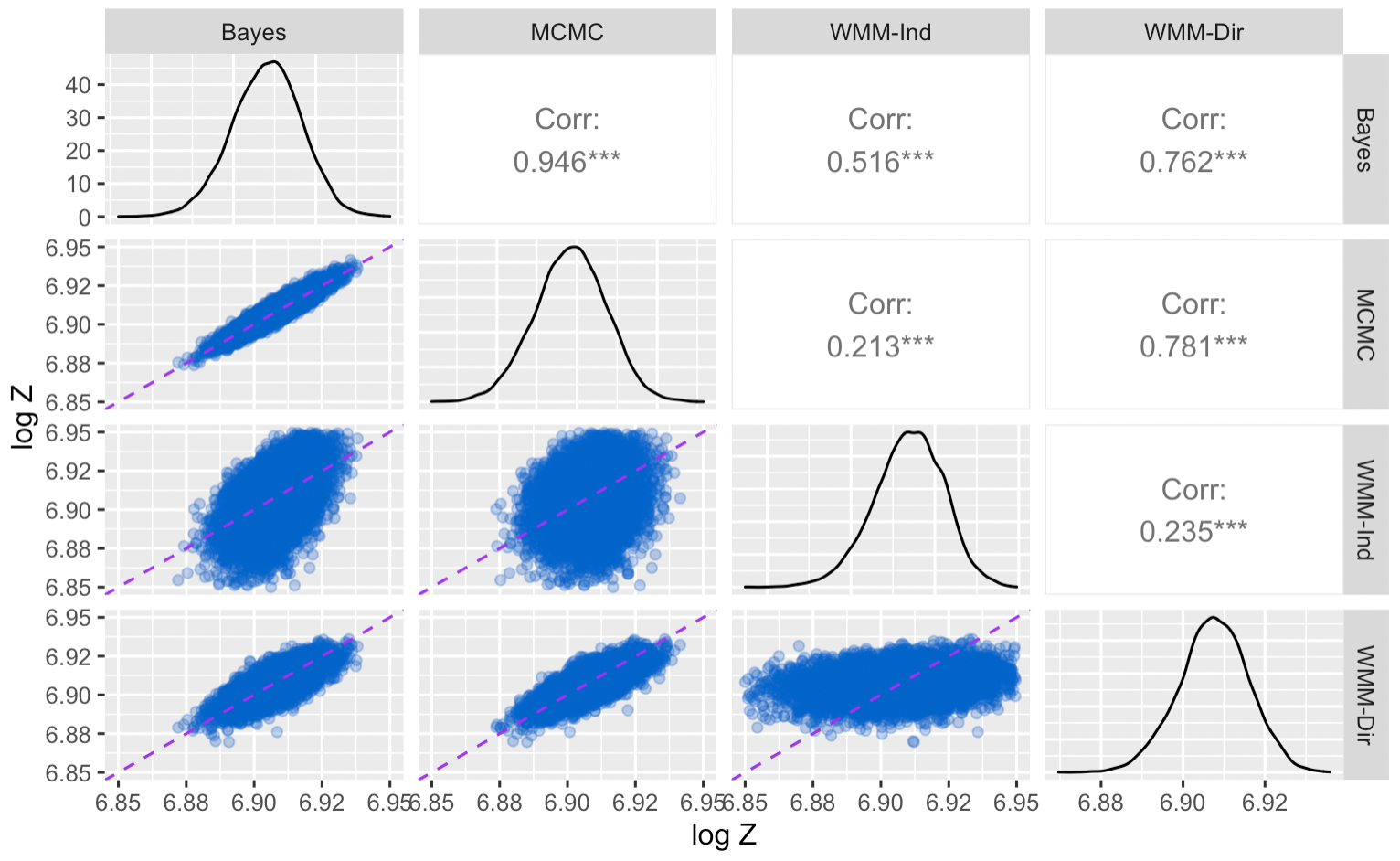}
	\caption[Simulation experiment 4 correlation plots comparing $\log Z$ estimates generated by closed-form Bayes solution, MCMC, and both methods of WMM.]{Experiment 4 - Correlation plots comparing $\log Z$ estimates generated by the closed-form Bayes solution, MCMC, and WMM methods in experiment 4, where prior $Z \sim Unif(0, 10000)$ and survey size $S=1000$.}
	\label{fig:corplotsexp2}
\end{figure}

\begin{figure}
	\centering
	\includegraphics[width=\linewidth,height=0.4\textheight]{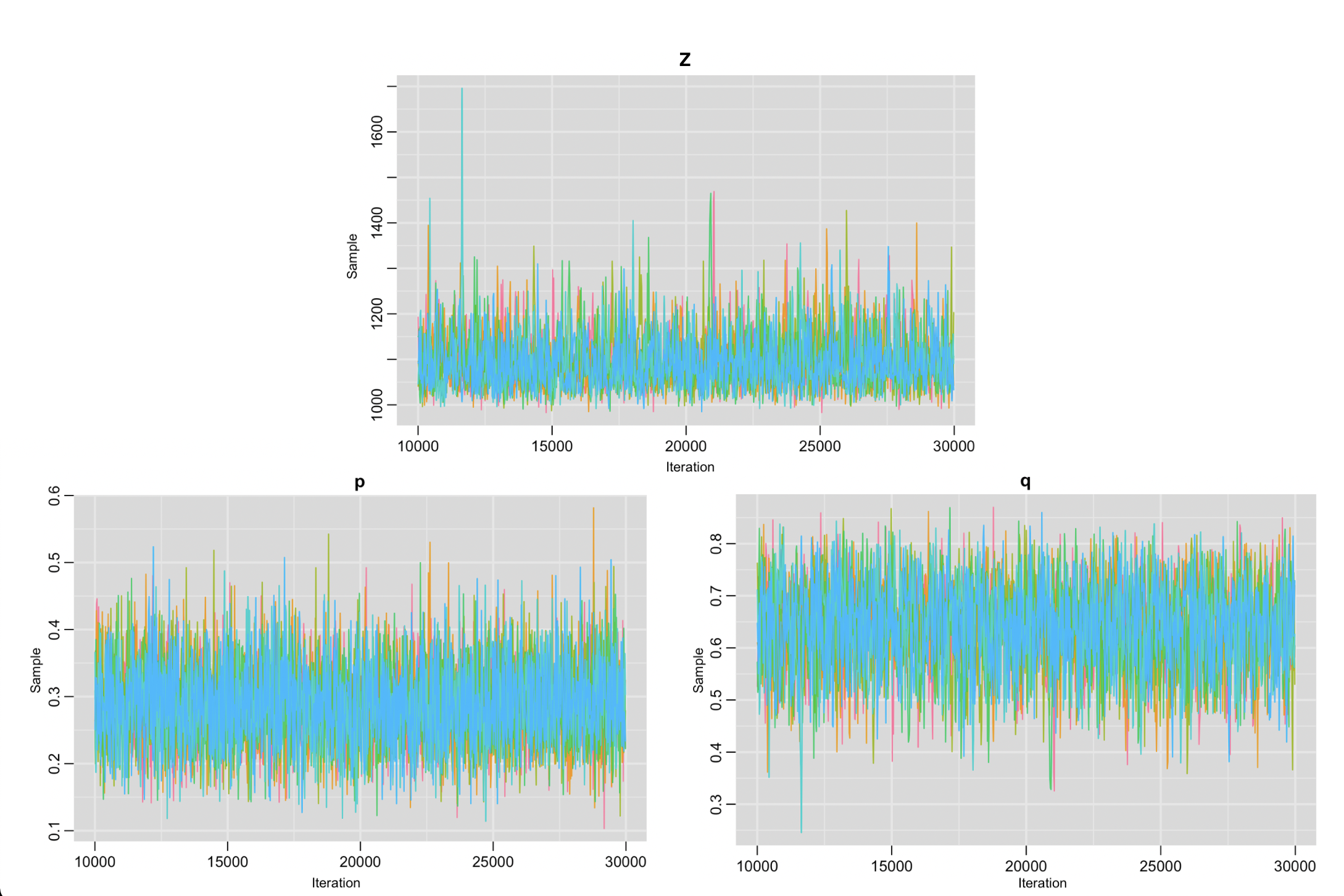}
	\caption{Experiment 5 - Trace plots of the MCMC output.}
	\label{fig:wrongpriortrace}
\end{figure}

\begin{figure}
	\centering
	\includegraphics[width=\linewidth,height=0.4\textheight]{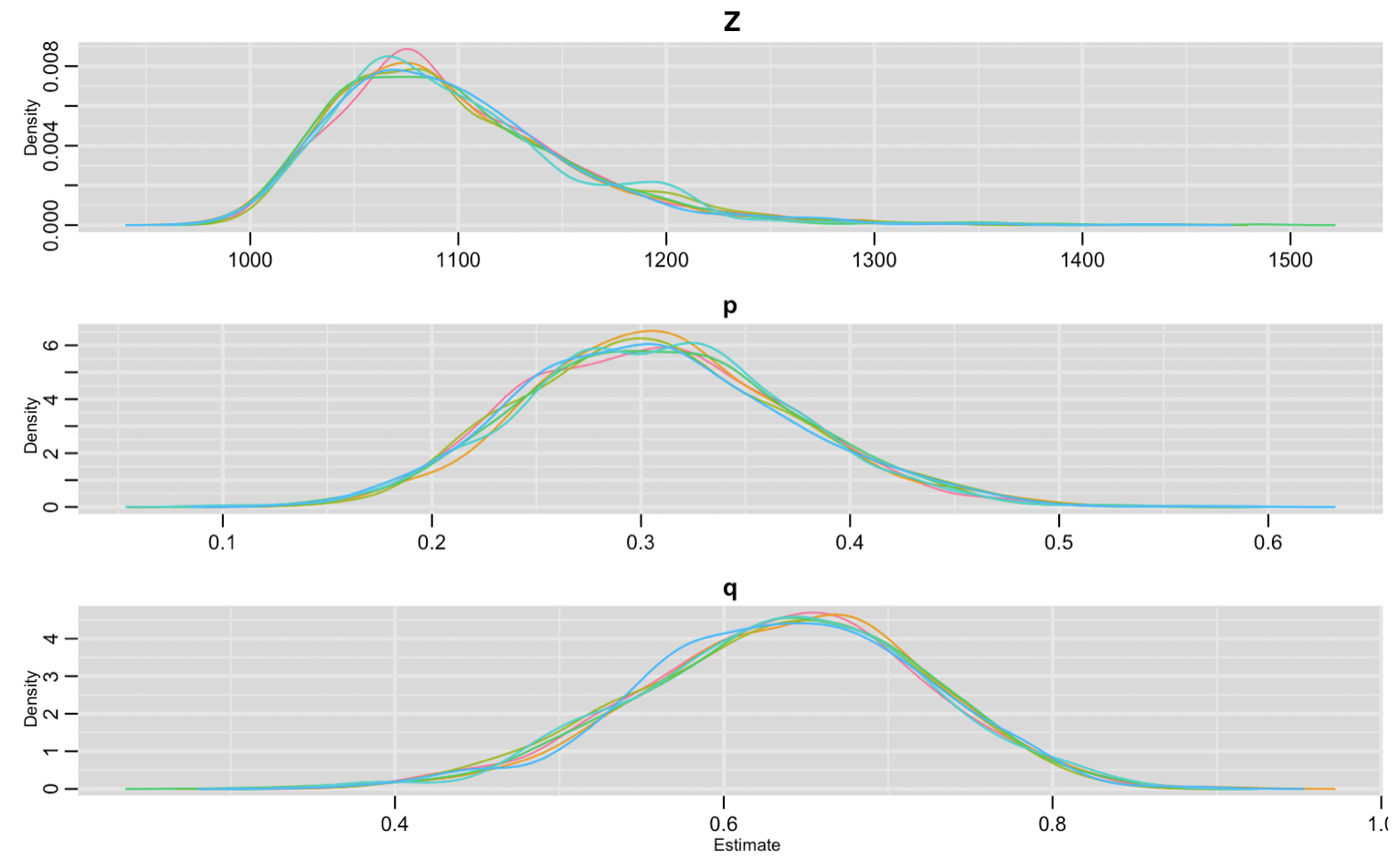}
	\caption{Experiment 5 - Density plots from the MCMC output.}
	\label{fig:wrongpriordens}
\end{figure}

\begin{figure}
	\centering
	\includegraphics[width=\linewidth,height=0.4\textheight]{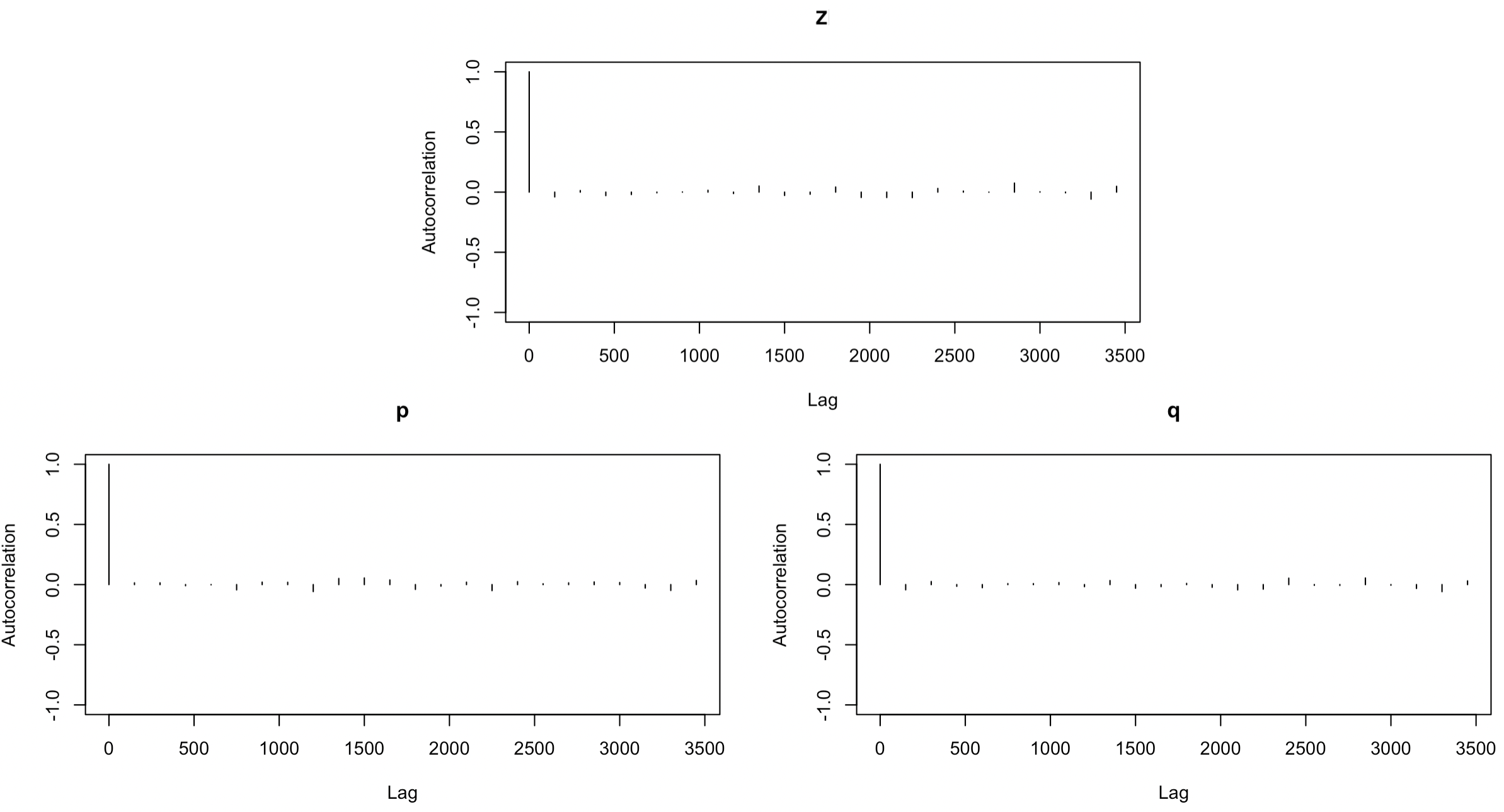}
	\caption{Experiment 5 - ACF plot from the MCMC output (example chain).}
	\label{fig:acfsim}
\end{figure}

\subsection*{Declaration of Interests}
The authors have no conflicts of interest to declare.

\subsection*{Acknowledgments}
This work was supported by NSERC Discovery Grant (RGPIN-2019-03957), an NSERC CGS-D, and a CIHR Doctoral Health System Impact Fellowship in partnership with the Public Health Agency of Canada (PHAC).

\clearpage

\bibliography{statrefs}


\end{document}